\documentclass[final,12pt]{elsarticle}
\usepackage[utf8]{inputenc}
\usepackage[T1]{fontenc}
\usepackage{fixltx2e}
\usepackage{graphicx}
\usepackage{longtable}
\usepackage{float}
\usepackage{wrapfig}
\usepackage{rotating}
\usepackage[normalem]{ulem}
\usepackage{amsmath}
\usepackage{textcomp}
\usepackage{marvosym}
\usepackage{wasysym}
\usepackage{amssymb}
\usepackage{hyperref}
\usepackage{amsmath}
\usepackage{pdflscape}
\usepackage{geometry}
\usepackage[multidot]{grffile}
\usepackage{todonotes}
\usepackage{subcaption}
\usepackage{wrapfig}
\usepackage{picinpar}

\begin{document}

\begin{frontmatter}

\title{PriMaL: A Privacy-Preserving Machine Learning Method for Event Detection in Distributed Sensor Networks}

\author[label1]{Stefano Bennati\corref{cor1}}
\ead{sbennati@ethz.ch}
\cortext[cor1]{Corresponding author}

\author[label2]{Catholijn M. Jonker}
\ead{C.M.Jonker@tudelft.nl}

\address[label1]{ETH Z\"{u}rich, Clausiusstrasse 50, 8092 Z\"{u}rich, CH}
\address[label2]{Interactive Intelligence, Fac. EEMCS, TU Delft, Mekelweg 4, 2628 CD Delft, NL}

\begin{abstract}
This paper introduces PriMaL, a general PRIvacy-preserving MAchine-Learning method for reducing the privacy cost of information transmitted through a network.
Distributed sensor networks are often used for automated classification and detection of abnormal events in high-stakes situations, e.g. fire in buildings, earthquakes, or crowd disasters.
Such networks might transmit privacy-sensitive information, e.g. GPS location of smartphones, which might be disclosed if the network is compromised.
Privacy concerns might slow down the adoption of the technology, in particular in the scenario of social sensing where participation is voluntary, thus solutions are needed which improve privacy without compromising on the event detection accuracy.

PriMaL is implemented as a machine-learning layer that works on top of an existing event detection algorithm.
Experiments are run in a general simulation framework, for several network topologies and parameter values.
The privacy footprint of state-of-the-art event detection algorithms is compared within the proposed framework.
Results show that PriMaL is able to reduce the privacy cost of a distributed event detection algorithm below that of the corresponding centralized algorithm, within the bounds of some assumptions about the protocol.
Moreover the performance of the distributed algorithm is not statistically worse than that of the centralized algorithm.

\end{abstract}


\end{frontmatter}

\section{Introduction}

Distributed sensor networks are applied to an ever-growing set of real-world domains \cite{rawat13_wirel_sensor_networ}, which is bound to grow even more with the further diffusion of Internet of Things (IoT) technologies \cite{whitmore2015internet,li2015internet}.
Sensor measurements in these networks are analyzed by an event detection algorithm and reported to some authority, referred to as supervisor.
Generally, a larger quantity of data obtained by the detection algorithm translates to a higher event detection accuracy.
The data measured by the sensors is often privacy-sensitive, particularly in the case of social sensing \cite{minson15_crowd_earth_early_warnin,pournaras16_self_regul_infor_sharin_partic_social_sensin} where sensors are related to specific people, and where this data can be used to infer user habits and preferences \cite{montjoye13_unique_crowd}.
The supervisor is modeled as a honest-but-curious adversary \cite{goldreich05_found_crypt_primer}, i.e. it performs the event detection accurately, but it can additionally run any type of algorithm on the data it receives, thus the more measurements are received the higher the potential privacy loss.

The tradeoff between privacy and accuracy is crucial in social sensing scenarios as participation in social sensing is usually voluntary, thus users could be motivated to leave the network if they perceive a high privacy cost.\\
\textbf{Case study: Detection of fires in households}\\
Consider a scenario in which each house is equipped with an infrared camera. These cameras can easily detect the start of a fire, but there are also many other events that might trigger a false alarm, for example an open oven.
The more contextual information is contained in the alarm, the more privacy sensitive information can be inferred by a honest-but-curious supervisor.
For example transmitting the light intensity or the infrared image of each room might help finding more accurately the source of fire, but might also reveal privacy sensitive information about the tenants.\\
\textbf{Case study: Earthquake detection through smartphones}\\
Consider a scenario in which an authority wants to promptly detect earthquakes in urban areas.
In order to do so, the authority asks the population to collaborate by sharing the measurements of their smartphone's movement sensors \cite{minson15_crowd_earth_early_warnin}.
Smartphones as an ensemble have the features of being broadly diffused, virtually covering all the territory, and being more concentrated in populated areas, where an earthquake can deal the most damage.
For the same reasons this solution has the disadvantage of being potentially very privacy-intruding, as every device is associated with a person thus it can be misused to infer the habits of the owner.
When dealing with a honest-but-curious supervisor, the mechanism has to ensure that reports cannot be linked back to a specific device.
Contextual information can help improving the accuracy but makes it easier for a malicious aggregator to link a series of reports to the same device  \cite{ma13_privac_vulner_publis_anony_mobil_traces}, thus it should be minimized.

Solutions are needed to increase user privacy without reducing detection accuracy.
Several privacy-enhancing techniques are proposed in the literature, for example anonymization techniques which prevent tracking users over time by breaking the connection between data and user identity, e.g. Mix-Zones \cite{beresford04_mix}.
A limitation of this solution is the need of a trusted middleware system that performs the anonymization.
In the scenario where third-party services cannot be trusted, users have to implement privacy-preserving techniques autonomously, from the bottom-up.
One such technique is obfuscation \cite{duckham05_formal_model_obfus_negot_locat_privac}, which degrades the quality of privacy-sensitive data, e.g. location, such that the exact value cannot be univocally determined.
Obfuscation has the disadvantage of introducing uncertainty in the measurement, which cannot always be tolerated, e.g. fire detection.

The goal of this paper is to develop a bottom-up method for reducing the privacy cost of communication in distributed sensor networks.
The increased sophistication and computational capabilities of IoT devices allows for equipping them with intelligent algorithms, which can be used to optimize individual properties of the devices.
Software optimization is preferable over ad-hoc design as the devices can be reprogrammed for different networks and protocols, as well as adapt to individual circumstances e.g. preferences of the owner.
Event detection algorithms in the literature use different flavors of machine learning to improve the detection accuracy, but to the best of our knowledge machine learning has not been applied yet to increasing privacy. 
It is assumed that each information transmitted over the network is associated to a privacy cost, so privacy can be increased by optimizing the contents of communication.

The contributions of this paper can be summarized as follows:
(I) Definition of a protocol and an algorithm for privacy-preserving event detection, which can be implemented on top of any existing detection algorithm.
(II) Definition of the criteria for a generally-applicable event detection algorithm.
(III) Implementation of PriMaL in a general framework and analysis of its performance and privacy footprint for several network topologies.
(IV) Application of PriMaL on several state-of-the-art detection algorithms, and comparison of its privacy footprint.
(V) Discussion of the effect of several parameter values on the privacy and accuracy of the event detection mechanism.

The rest of the paper is organized as follows:
Section \ref{sec:litrev} contains a description of state-of-the-art event detection algorithms, and of criteria for the general applicability of an event detection algorithm.
Section \ref{sec:model} describes the setting formally.
Section \ref{sec:method} describes in detail the methodology and the concepts behind it. Section \ref{sec:results} presents the results of the simulations and discuss their implication.
Section \ref{sec:rel_work} analyzes and compares the privacy footprint of several state-of-the-art algorithms.
Section \ref{sec:discussion} concludes with a discussion of the findings and indications for future developments.


\section{Background Literature}
\label{sec:litrev}
Three types of actors are common in the literature: sensors, agents and supervisors.
\emph{Sensors} are hardware devices that perceive the environment and output a measurement at regular intervals. It is assumed that their specifications are given and their behavior is fixed.
\emph{Agents} are entities equipped with some machine-learning algorithm, e.g. a classifier. They cannot measure directly but have the ability to process and classify measurements. It is assumed that they can run arbitrary algorithms.
\emph{Supervisors} are physical or virtual entities that are connected to every agent and log alarms reported by them.
Supervisors can additionally provide agents with feedback about the classification accuracy, or perform some operation on the alarms, typically computing the consensus between individuals \cite{bahrepour09_sensor_fusion_event_detec_wirel_sensor_networ,bahrepour10_distr_event_detec_wirel_sensor}.
The goal of the network is to classify events accurately and inform the supervisor about this classification.

Sensor measurements are commonly assumed to be correlated in space or time
\cite{zhang09_adapt_onlin_one_class_suppor,faulkner13,ruan08_binar,bahrepour09_sensor_fusion_event_detec_wirel_sensor_networ,bahrepour10_distr_event_detec_wirel_sensor,zhang12_statis_based_outlier_detec_wirel_sensor_networ}. This work assumes instead that agents which process the same event type can communicate with each other.
Each agent is equipped with a classifier, trained on the local measurements  \cite{faulkner11,zhang09_adapt_onlin_one_class_suppor,ruan08_binar,bahrepour10_distr_event_detec_wirel_sensor}.
This has the advantage over solutions with a single classifier or global threshold (e.g. \cite{faulkner13,bahrepour09_sensor_fusion_event_detec_wirel_sensor_networ}) that the system can capture local differences in the sensor readings, e.g. environmental conditions. The disadvantage of this approach is that accuracy is lower because of the reduced size of training data for each sensor \cite{wittenburg10}.

Network organizations commonly found in the literature can be classified in centralized, decentralized and distributed.
In a \emph{centralized} organization a single agent is in charge of the whole sensor network: it receives the information from all sensors and decides individually whether to send an alarm.
The communication footprint of a centralized system is constant and proportional to the number of sensors.
Receiving all information in the network allows the agent to take optimal decisions, but the agent becomes a single points of failure that undermines the robustness and reliability of the system.
In a \emph{decentralized} system, each sensor is equipped with a local algorithm that decides whether to send an alarm based on local experience.
A decentralized system transmits less measurements when compared to a centralized organization, as decisions are taken locally so the majority of measurements are not transmitted over the network.
As opposed to a centralized organization, this configuration does not have any single points of failure, so it is more resilient to malfunctioning and failure.
The disadvantage is that each agent acts independently of the others, so a voting algorithm is often required to aggregate the individual decisions and come to a consensus.
A \emph{distributed} system is similar to a decentralized system, with the difference that sensors can communicate with each other in order to share experience \cite{zhang09_adapt_onlin_one_class_suppor}, or to take collective decisions \cite{ruan08_binar,bahrepour10_distr_event_detec_wirel_sensor}.
In this setting communication footprint is high and it is proportional to the connectivity between agents.

Communication is limited to the 1-hop neighborhood of agents, to avoid routing issues and reduce power consumption \cite{wittenburg10}.
Messages can contain measurements \cite{bahrepour10_distr_event_detec_wirel_sensor,zhang12_statis_based_outlier_detec_wirel_sensor_networ}, individual classifications labels  \cite{zhang09_adapt_onlin_one_class_suppor,ruan08_binar,bahrepour10_distr_event_detec_wirel_sensor,zhang12_statis_based_outlier_detec_wirel_sensor_networ,marin-perianu07_d_fler_distr_fuzzy_logic}, classification parameters \cite{zhang09_adapt_onlin_one_class_suppor,wittenburg10}, or reputation scores \cite{bahrepour10_distr_event_detec_wirel_sensor}. In this work the content of messages is defined by the protocol described in Section \ref{sec:model}.
A way to limit the computational and networking requirements is to transmit only messages classified as outliers
\cite{zhang09_adapt_onlin_one_class_suppor,zhang12_statis_based_outlier_detec_wirel_sensor_networ,faulkner11}.

Outlier classification can be performed using a variety of techniques \cite{shahid12_charac_class_outlier_detec_techn}: statistical methods classify measurements based on their underlying probability distribution, of which a model is required a priori. Clustering algorithms use machine learning techniques to group measurements in different classes, they require the definition of a distance measure. Classification algorithms detect outliers by drawing a boundary that separates normal measurements and events, these algorithms do not require a model of the data but require training, either supervised or unsupervised. Examples of classification algorithms are Support Vector Machines, which use a non-linear function (kernel) to map the input into a higher-dimensional space and Bayesian Networks which generate a graph of random variables, whose connection represent conditional probabilities.
Training of these algorithms can be done online \cite{zhang09_adapt_onlin_one_class_suppor,ruan08_binar,ruan08_binar} or offline during a training phase \cite{wittenburg10,bahrepour09_sensor_fusion_event_detec_wirel_sensor_networ}.
Clustering and classification are the most popular techniques, as they can be used in both decentralized and distributed systems. The approach used in this work is classification, specifically One-class SVMs \cite{laskov04_intrus_detec_unlab_data_with}, which produces a binary classification.
Nevertheless, the mechanism described in this paper does not depend on a specific classification algorithm.
Other approaches support a n-ary classification in the number of event types \cite{wittenburg10}.
Detection might be noisy i.e. flipped with a certain probability \cite{faulkner13}. In this work the network is assumed to be reliable, so detection labels are always correct. Relaxing this assumption and dealing with sensor failures is left to future work.

The performance of different algorithms is measured with the standard measures of Precision and Recall.

\subsection{Criteria for a generally-applicable detection algorithm}
A generally-applicable algorithm should run on any network organization, therefore measurement and detection should be separated \cite{marin-perianu07_d_fler_distr_fuzzy_logic}.
This separation allows for assigning an arbitrary number of sensors to an agent running an event detection algorithm, a prerequisite for supporting both a centralized a organization, where all sensors are connected to a single agent, and a decentralized organization, where there are as many agents as sensors.
The detection algorithm used in the paper is based on the distributed implementation of \cite{zhang09_adapt_onlin_one_class_suppor}, as it is parsimonious in communication and satisfies the following requirements:

\begin{itemize}
\item It supports networks with heterogeneous sensors. Heterogeneous networks can be seen as a combination of homogeneous subnetworks, where every type of sensor is processed separately. Two sensors are of different type if either they measure different event types, or if they measure the same event type but output a different signal, e.g. measure fire with smoke detectors or heat detectors.
\item Each agent has its own classification algorithm which allows each sensor to learn independently of the others, thus supporting a distributed organization.
\item Measurements are not assumed to be correlated in space or time.
\end{itemize}

In order to model a distributed organization, agents are allowed to exchange their measurements with their 1-hop neighborhood and receive their neighbors' opinion.
The system is robust to failures as every agent is able to produce alarms even if there are no neighbors to communicate with.
Another advantage of independent classifiers is that every agent could have one classifier for each sensor it is connected to. This makes the system more resilient to malfunctioning sensors, making the system suitable for networks with cheap and unreliable hardware. Moreover each classifier can adapt to its local conditions, for example locations with high background noise.

All these requirements ensure the general applicability of the method, moreover the independence of this method from a specific classifier allows it to be trained either offline or online.


\section{Problem Formulation}
\label{sec:model}

The environment is a set of locations $\mathcal{L}$. A signal $x_l^t$, the ground truth, is generated randomly by some sensor at each location $l$ and each time $t$, by sampling from some unknown random variable $X_l$.
A threshold function $\tau_l$ determines whether a signal $x_l^t$ is an event or not: it returns \emph{true} if the signal is an event, \emph{false} otherwise.
No assumptions are made about this function, e.g. it can be a human supervisor.
$\mathcal{S}$ is the set of sensors in the environment and $L_s$ is the set of locations measured by the sensor $s$.
By assumption, each location is covered by exactly one sensor.
$\mathcal{A}$ is the set of agents, each of which is connected to a set of sensors $S_a \subseteq \mathcal{S}$. 
No overlap is allowed between these sets: $S_a,S_{a^\prime} \subseteq \mathcal{S}, a \ne a^\prime \in \mathcal{A} \Rightarrow S_a \cap S_{a^\prime} = \emptyset$.
$\forall s \in \mathcal{S}, ~ \tau^\prime_s$ is the optimal classification function for sensor $s$, the value that minimizes false positives and negatives.
The function $\tau^\prime_s$ is learned by the agent responsible for sensor $s$ and outputs a discrete value and confidence interval, for each of its sensors $s \in S_a$, which means that for every location $l$ a corresponding $\tau^\prime_l$ is learned.
Each agent can communicate with other agents by emitting a message $m^t_l$.
A message $m^t_l \subseteq \langle a_{ID},l,x_l^t,t,e_{TYPE},s_{TYPE} \rangle$ is composed by:
\begin{itemize}
\item $a_{ID}$  the agent ID.
\item $l$ the location where the event was measured (corresponds to the sensor ID).
\item $x_l^t$ the value of the signal as measured by the sensor at that location.
\item $t$ the timestep.
\item $e_{TYPE}$ the type of event.
\item $s_{TYPE}$ the type of sensor that measured the event.
\end{itemize}
Messages can be sent to ask for an opinion from other agents. As a response, agents emit messages indicating their classification for transmitted event, which can take values of \emph{true}, \emph{false}, or \emph{unknown} if the event type is not valid or if the confidence about the classification is low.
Similarly agents can communicate to the supervisor, in order to report events, by triggering alarms $w_l^t$ with the same structure as normal messages.
All data that is transmitted over the network is assumed to be subject to a transmission cost, except for data transmitted locally from sensor to agent, which has neither communication nor privacy cost.
All communication over the network is assumed to be eavesdropped, thus it has a privacy cost if it contains sensitive information, e.g. the sensor ID.
It is assumed that the sensor ID cannot be inferred when intercepting a message from an agent, even if that agent is connected to only one sensor. This assumption is reasonable for systems that randomly generate agent IDs at every transmission, e.g. mix zones \cite{beresford04_mix}.

Performance is evaluated according to the following criteria:
\begin{itemize}
\item Accuracy is defined as follows:
\begin{itemize}
\item True Positive (TP) if $\tau_l(x^t_l)=\mbox{true}$ and $w_l^t$ has been triggered.
\item False Positive (FP) if $\tau_l(x^t_l)=\mbox{false}$ and $w_l^t$ has been triggered.
\item True Negative (TN) if $\tau_l(x^t_l)=\mbox{false}$ and $w_l^t$ has not been triggered.
\item False Negative (FN) if $\tau_l(x^t_l)=\mbox{true}$ and $w_l^t$ has not been triggered.
\item Precision: TP/(TP+FP)
\item Recall: TP/(TP+FN)
\item F-Measure: 2*(Precision*Recall)/(Precision+Recall)
\end{itemize}
\item Communication cost: It depends on the application, e.g. proportional to bandwidth used or power consumption.
\item Privacy cost: It depends on the application, e.g. the quantity of privacy-sensitive pieces of information contained in the measurement.
\end{itemize}


\section{Experimental Setting}
\label{sec:method}
This section describes technical details about the framework and introduces the experimental setup.

\subsection{Framework}

The event detection process of a network of sensors is simulated as follows:

\begin{enumerate}
\item A ground truth for the event is randomly generated, indicating what sensor perceives an event and at what time.
\item A measurement is generated for each sensor, whose value is sampled, depending on the ground truth, from one of two probability distributions, one determining the events and the other the normal measurements.
\item Each sensor performs a measurement and reports it to the agent to which it is connected.
\item Each agent receives one or more measurements and performs classification on them.
\item Optionally and depending on the network organization, agents exchange measurements via the network.
\item Each agent reports alarms to the supervisor.
\item The supervisor logs all alarms reported by the network and, optionally, provides feedback about the classification that can be used to train classifiers.
\end{enumerate}

\begin{figure}[t]
  \centering
  \includegraphics[width=0.9\textwidth]{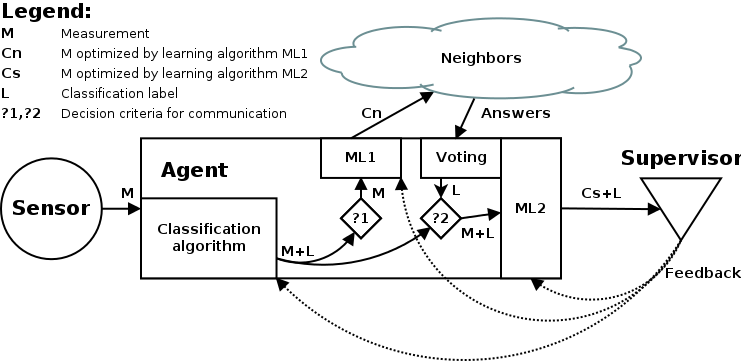}
  \caption{Flowchart describing the operations of the framework.}
  \label{fig:flowchart}
\end{figure}

Figure \ref{fig:flowchart} describes visually how the framework operates, from the point of view of an agent.

Each agent classifies all events it receives with a One-Class SVM classifier.
Each classifier is trained on the measurements corresponding to only one specific event type, and is used to classify only events of the same type.
Alternatively agents could train a different classifier for each sensor.
The advantage of the latter solution is that classifiers can adapt better to individual conditions, the disadvantage is that each classifiers receives less measurements, which can slow down online learning.

Neighbors can help an agent with the classification by providing their opinion on measurements received by the agent.
Each agent combines the opinion of the neighbors by majority voting, where each vote is weighted by its classification confidence.
Combining the experience of agents can increase detection accuracy but increases the communication on the network.
If the own classification confidence is high, the opinions of the neighbors should not change the decision, in this case it is desirable to not transmit the message, in order to reduce the communication footprint.
If the measurement is eventually classified as an event, the agent sends a signal to the supervisor. The feedback potentially given by the supervisor can be used to train the classifiers.

Each piece of information contained in a message is assumed to have a strictly-positive communication cost and a non-negative privacy cost, so reducing the amount of information transmitted reduces the costs.
There can be situations where some piece of information has a very high privacy cost, e.g. gps location, and the information it conveys can be approximated by using a combination of other pieces of information that have a lower privacy cost, e.g. list of visible Wi-Fi networks. In this case substituting the first piece of information for the other pieces increases the communication cost but reduces the privacy cost.
Moreover, the transmission of some information might not be required to achieve the goal of the communication, for example the ID of the sensor reporting a measurement is not needed for its classification.
This system works under the assumption that the communication is subsettable: it is composed by several independent pieces of information that are meaningful even when isolated. This is the case for the protocol defined in Section \ref{sec:model}.

A learning algorithm can be used to optimize the content of messages, by learning the cost in terms of communication and privacy of each piece of information.
As the information required by the supervisor and by the neighbors might differ, the agents are equipped with two separate learning algorithms that optimize independently the two communication channels.
The machine learning algorithms need training. If training is executed online, in a phase denoted as \emph{calibration}, the algorithm learns by trying different combinations and recording the cost of communication. If some fundamental piece of information is omitted, e.g. the event location in an alarm, the communication fails and a high cost is reported to the learning algorithm. These failures might also reduce the event detection accuracy.

\subsection{Experimental Setup}

A network organization is defined by the connections between agents and sensors.
The total cost of communication varies with the organization, each of which has a specific communication pattern, and with parameter settings such as number of sensors and size of the neighborhood.
All experiments are run on specific network organizations:

\begin{figure}[t!]
  \centering
  \begin{minipage}[t]{0.4\linewidth}
  \centering
  \includegraphics[height=0.2\textheight]{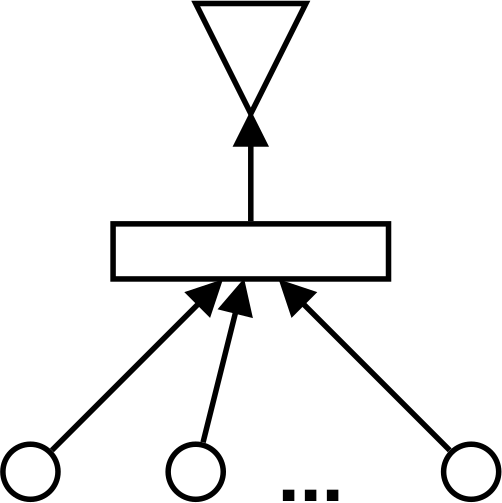}
\subcaption{Centralized.}\label{fig:topo_cen}
  \end{minipage}
  \begin{minipage}[t]{0.4\linewidth}
  \centering
  \includegraphics[height=0.2\textheight]{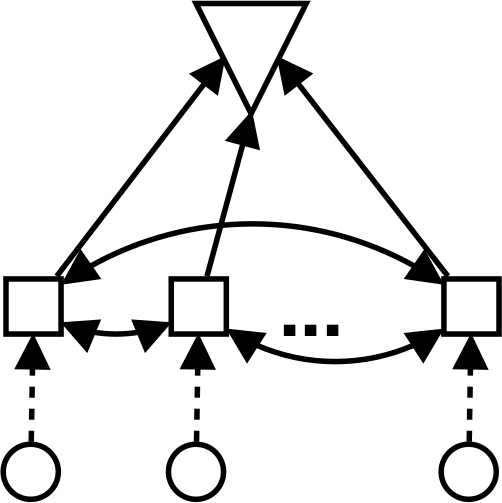}
\subcaption{Fully distributed.}\label{fig:topo_distr}
\end{minipage}
  \begin{minipage}[t]{0.15\linewidth}
  \includegraphics[height=0.2\textheight]{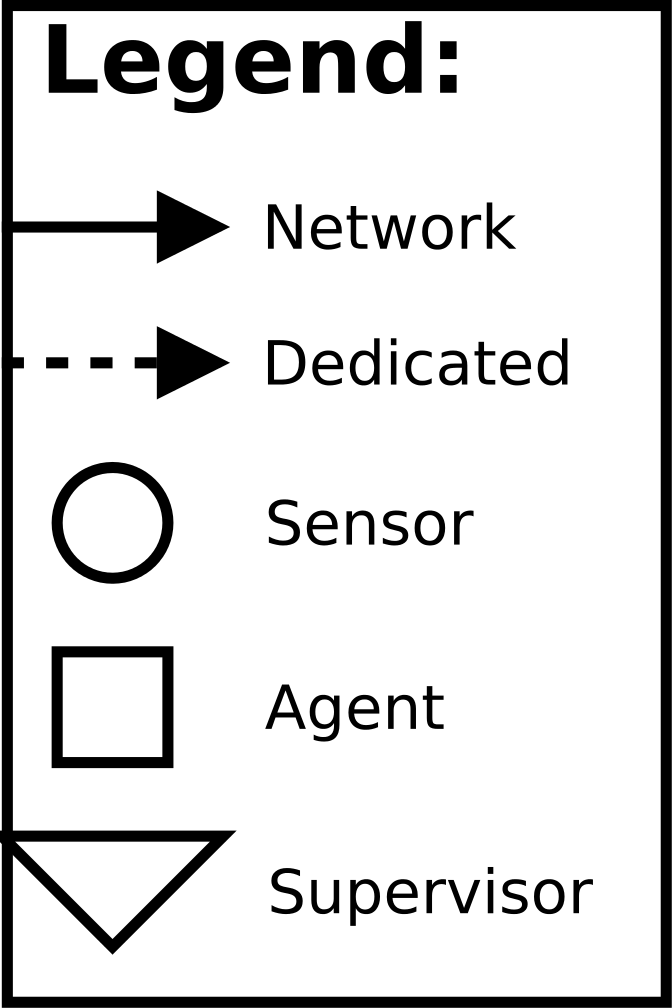}
  \end{minipage}
  \begin{minipage}[t]{0.9\linewidth}
  \centering
  \includegraphics[height=0.2\textheight]{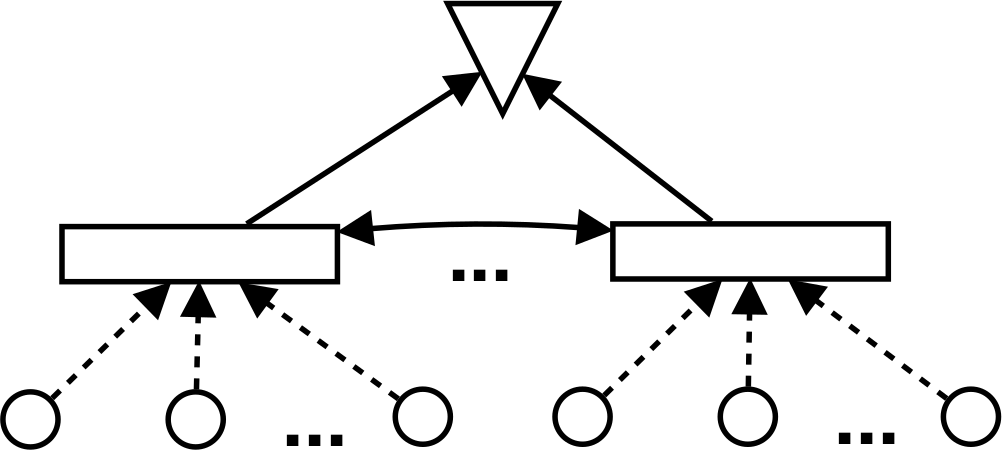}
\subcaption{Partially distributed.}\label{fig:topo_part}
  \end{minipage}
  \caption{Illustrations of different organizations.}
  \label{fig:topo1}
\end{figure}

In a \emph{centralized} system, only one agent is connected to all sensors (Figure \ref{fig:topo_cen}).
$\mathcal{A}=\{a\}, S_a = \mathcal{S}$.
All sensors transmit over the network so the transmission and privacy costs are equivalent and constant.
In a \emph{decentralized} system every agent is connected to exactly one sensor (Figure \ref{fig:topo_distr}).
$S_a,S_{a^\prime} \subseteq \mathcal{S}, a \ne a^\prime \Rightarrow S_a \cap S_{a^\prime} = \emptyset$.
Every agent trains an independent classifiers for its own sensor.
Sensors transmit to the agent through a dedicated channel, as they are physically connected, thus they do not pay any cost of communication.
Agents instead communicate over the network, so they pay a cost to transmit to the supervisor.
A \emph{distributed} system is similar to a decentralized system, but additionally agents can communicate with each other over the network.
At every timestep, the agent can send a message to its neighbors and receive as an answer the output of the classification functions of the neighbors.
The agent then produces a new classification for the event by aggregating its own classification and the neighbor's classification by majority vote.

Experiments are run within the framework described above in order to evaluate the proposed mechanism.
The main feature of PriMaL is the learning algorithm that filters the information to be transmitted.
Learning algorithms can be trained either offline, with some historical data, or online, with data collected by the system in operations.
If the latter training method is chosen, the system goes trough an initial calibration phase where the learning algorithm collects experience and might not perform as expected.
In the scenario outlined in this paper, the learning algorithm operates on the content of communication, so during the calibration phase messages might not contain the information they are supposed to. If this is the case, the detection accuracy can decrease.
The tradeoff between privacy and accuracy during calibration needs to be explored, in order to confirm the validity of the method, and this is done over several experiments.

The \textbf{first experiment} explores the effect of calibration on performance where a fully distributed network (cf. Figure \ref{fig:topo_distr}) is equipped with uncalibrated learning algorithms and compared to a network where the learning algorithm is disabled.

The \textbf{second experiment} compares privacy and accuracy of a detection algorithm for varying parameter values and network organizations.
Parameters such as the number of sensors and the size of the neighborhood have an influence on the communication footprint of the algorithms (cf. Section \ref{sec:rel_work}), thus on its privacy footprint and accuracy.
In order to compare different simulations, the measures are taken on a system that is already calibrated and that reached a steady state.

The choice of what to communicate is as important as the accuracy of classification for the performance of the system.
So far it has been assumed that the two correspond: only events with a positive classification are transmitted.
The \textbf{third experiment} relaxes this assumption and compares the effect of different criteria on privacy and accuracy.
Classification has two characteristics: the label and the confidence.
A transmission criterion based on the classification label is to transmit only measurements with positive classification. This criterion works well if classification is confident, but it might lead to bad performance if the confidence is low.
A criterion based on the confidence is to transmit all measurements for which the confidence is low. This criterion is suitable for improving the average classification confidence as it can be used to ask for the opinion of neighbors in a distributed organisation.
All simulations are run with the same parameters in a fully distributed organization (cf. Figure \ref{fig:topo_distr}), and are compared to a baseline with classification disabled, which sends all messages to both the neighbors and the supervisor.
Note that all combination that involve transmitting all measurements to the supervisor have the same performance as the baseline, thus they are omitted from the comparison.


\section{Results}
\label{sec:results}

\begin{figure}[t!]
  \centering
  \begin{minipage}[t]{0.45\linewidth}
  \includegraphics[width=\linewidth]{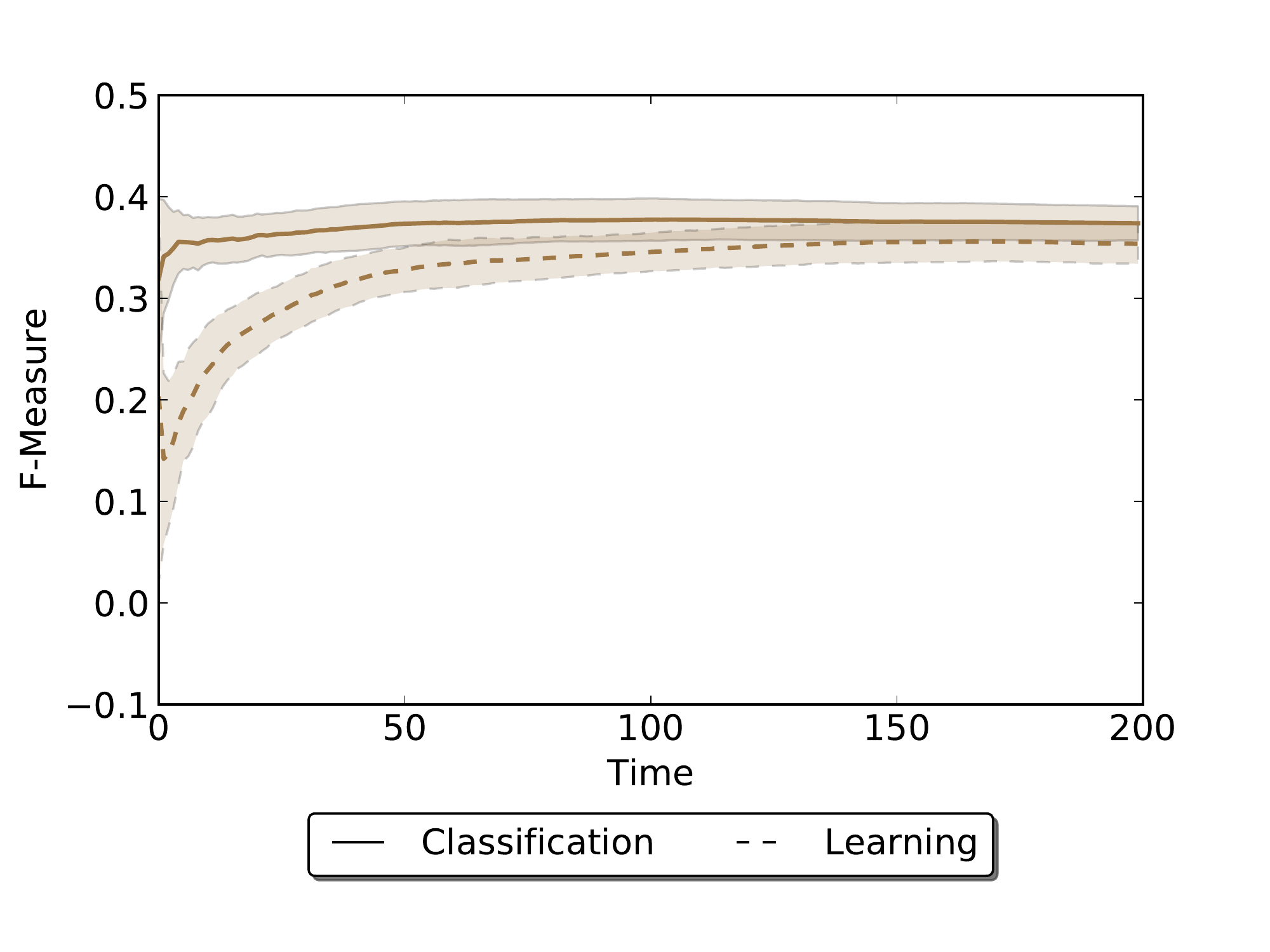}
\subcaption{Classification accuracy}\label{fig:learn2l}
  \end{minipage}
  \begin{minipage}[t]{0.45\linewidth}
     \includegraphics[width=\linewidth]{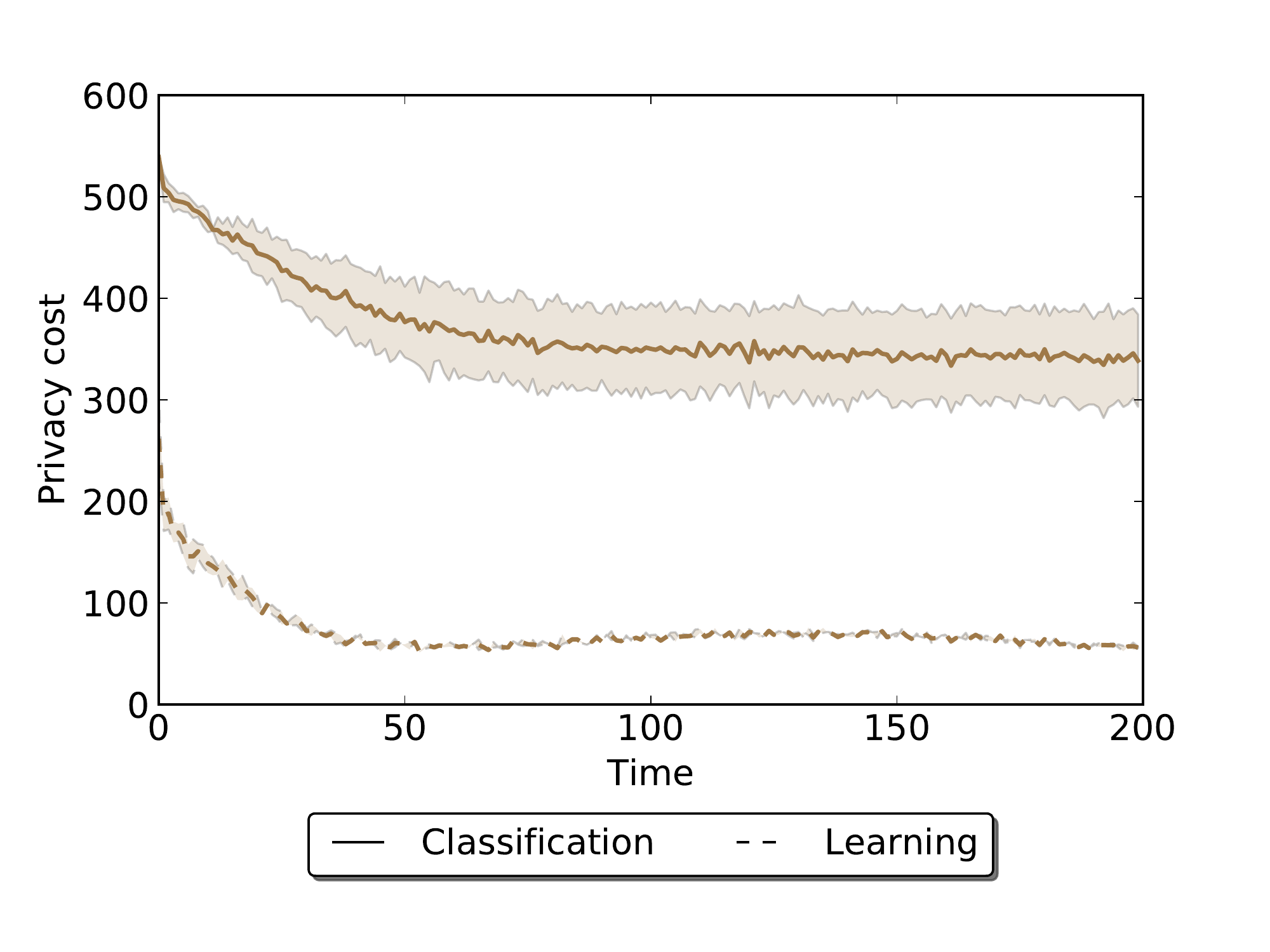}
\subcaption{Privacy cost}\label{fig:learn2r}
  \end{minipage}
  \caption{First experiment. Effect of learning calibration on classification. Shaded area corresponds to 0.95 confidence interval. }
  \label{fig:learn2}
\end{figure}

The \textbf{first experiment} investigates the effect of calibration on privacy and detection accuracy.
Figure \ref{fig:learn2l} shows that a network with learning agents (dashed line) starts at a lower performance level than a network without learning agents (continuous line), but the performance of the former grows until it reaches that of latter.
Figure \ref{fig:learn2r} shows that this reduction in performance is compensated by a reduction in privacy cost starting very early in the simulation. In other words, the learning algorithm reduces both the performance and the privacy cost during the calibration phase, but performance recovers quickly to a normal value.
In both cases the classifiers learn online, therefore accuracy improves over time and the number of messages transmitted, thus the privacy cost, decreases e.g. false positive reports, requests to the neighbors.
Important to note that the learning algorithm of choice is standard and not optimized, so the calibration time could be substantially reduced by a more sophisticated algorithm.

\begin{figure}[t!]
  \centering
  \begin{minipage}[t]{0.45\linewidth}
\includegraphics[width=\linewidth]{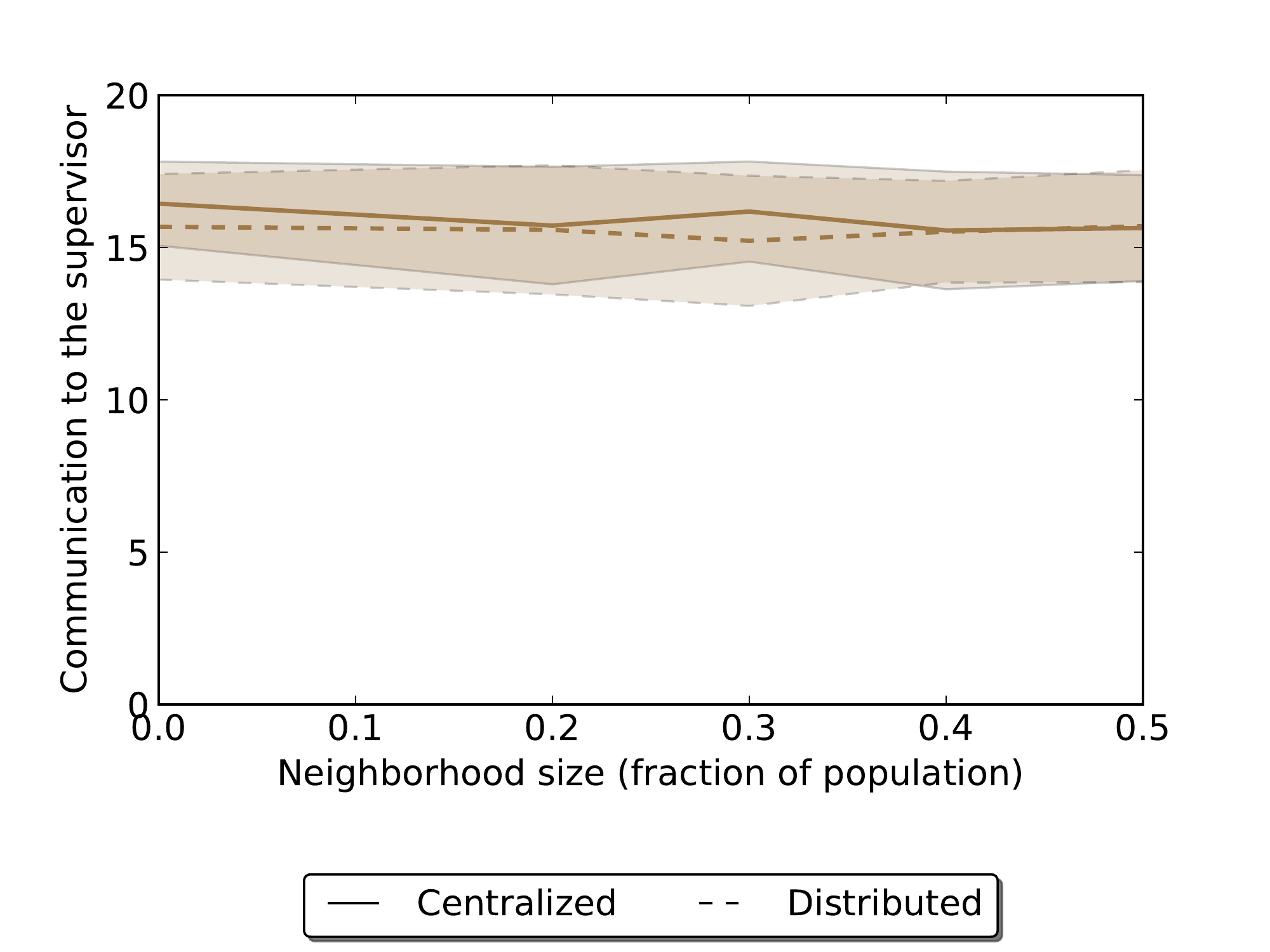}
  \end{minipage}
  \begin{minipage}[t]{0.45\linewidth}
    \includegraphics[width=\linewidth]{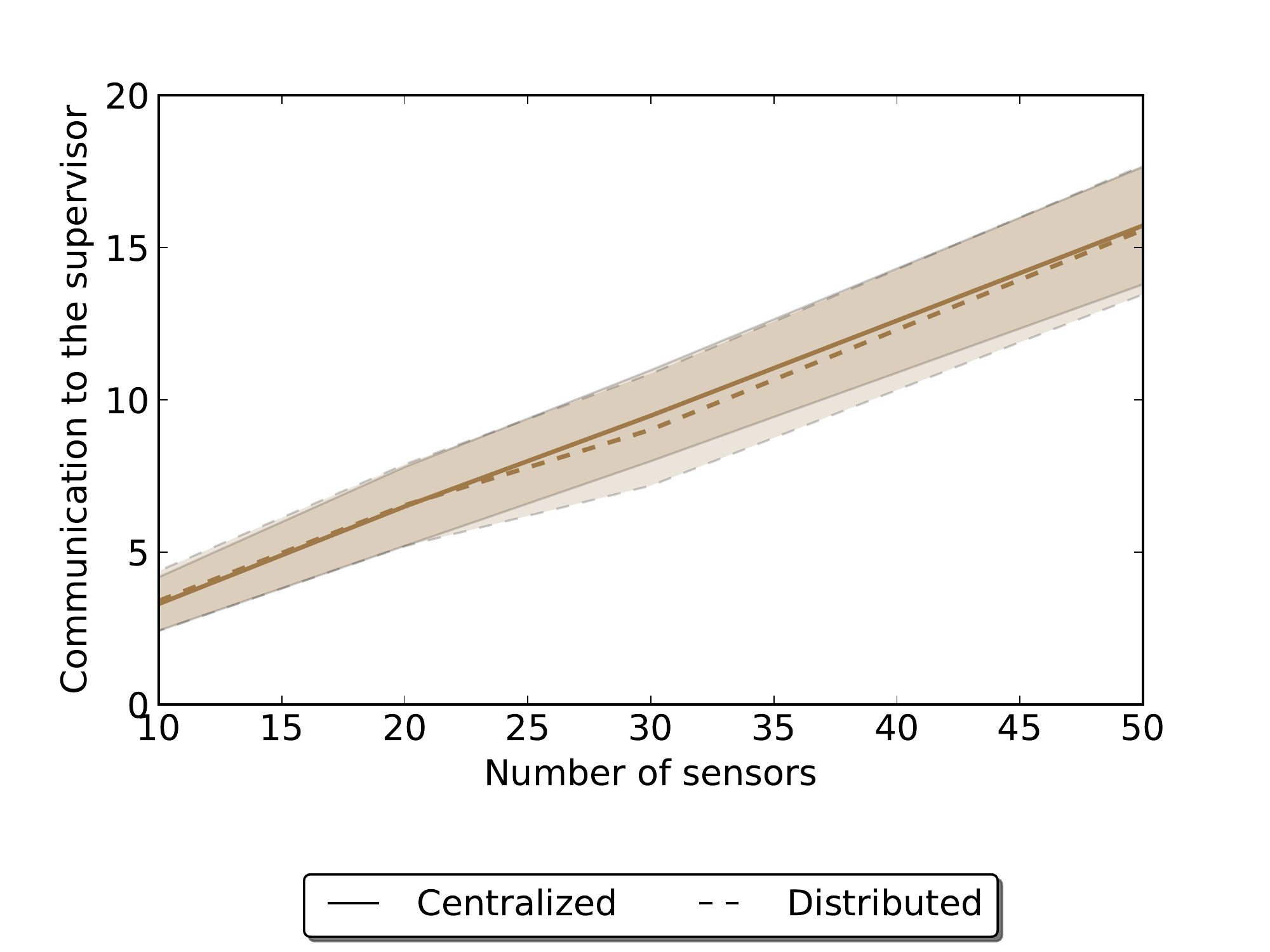}
  \end{minipage}
  \caption{Second experiment. Communication frequency towards the supervisor for different parameter configurations. A centralized and a distributed organization produce the same results if classification is disabled. Shaded area corresponds to 0.95 confidence interval.}
  \label{fig:param1}
\end{figure}

\begin{figure}[t!]
  \centering
  \begin{minipage}[t]{0.45\linewidth}
\includegraphics[width=\linewidth]{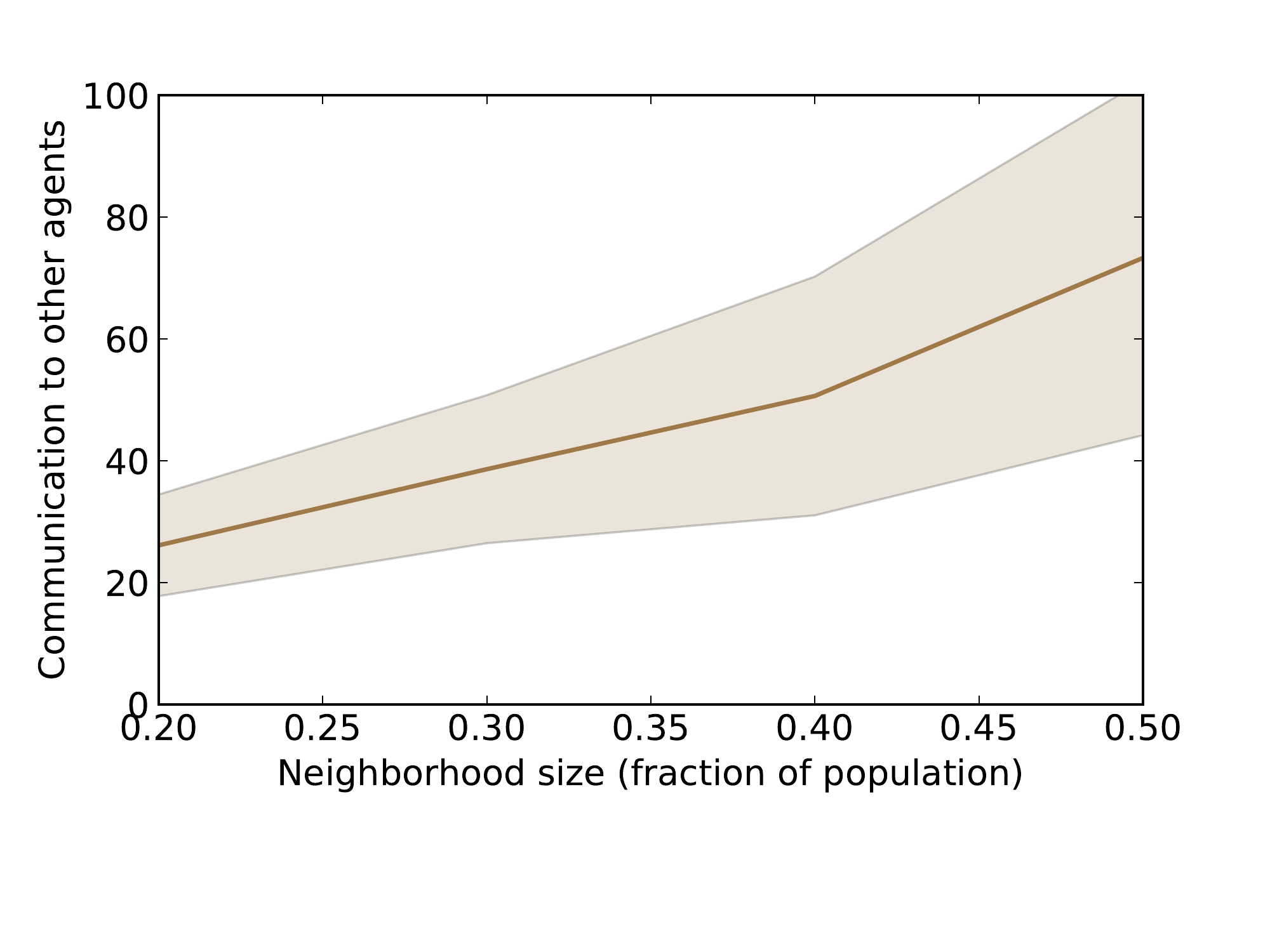}
  \end{minipage}
  \begin{minipage}[t]{0.45\linewidth}
\includegraphics[width=\linewidth]{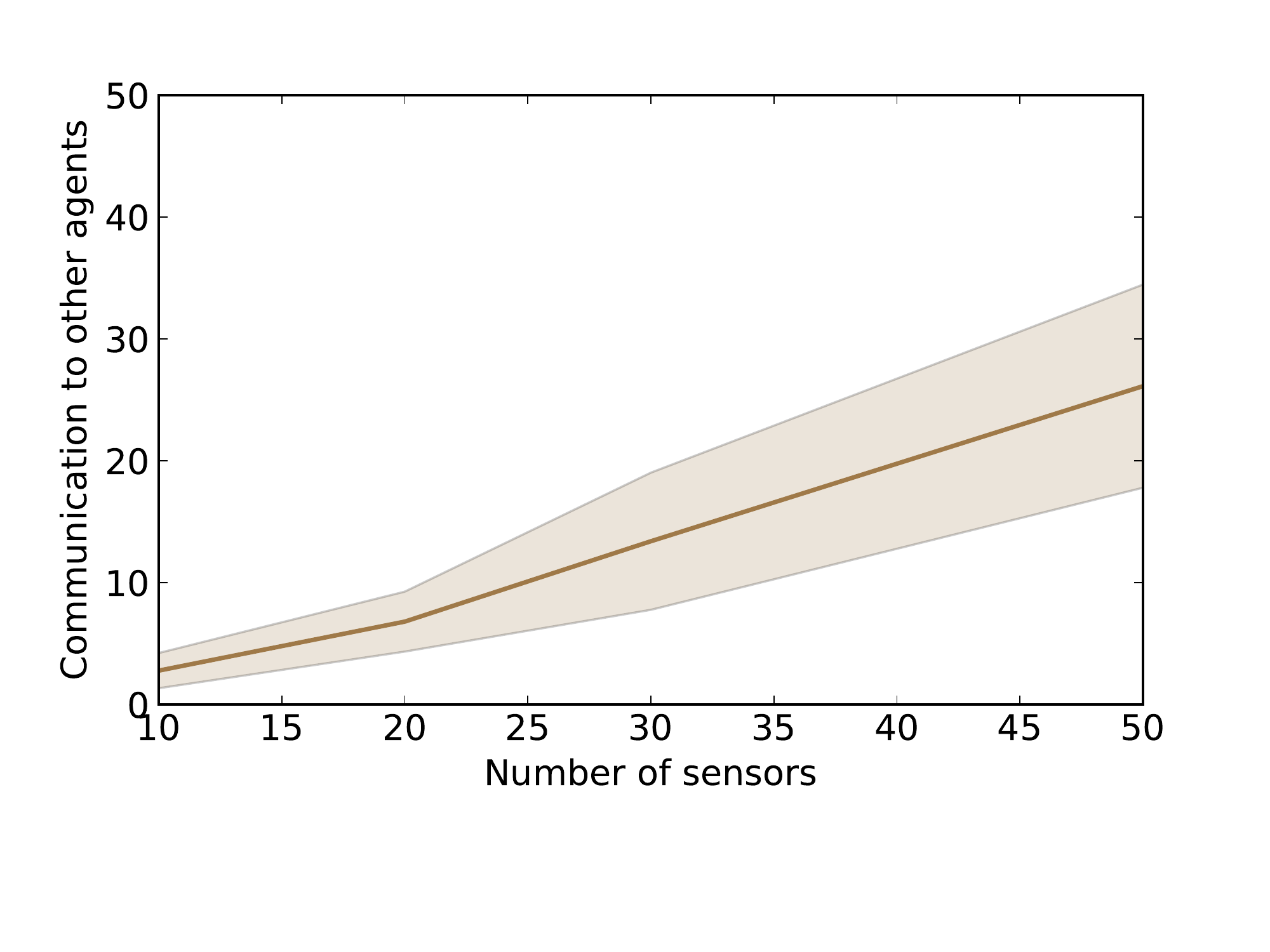}
  \end{minipage}
  \caption{Second experiment.  Communication frequency towards neighbors for a distributed organization. Shaded area corresponds to 0.95 confidence interval.}
  \label{fig:param2}
\end{figure}

\begin{figure}[t!]
  \centering
  \begin{minipage}[t]{0.45\linewidth}
\includegraphics[width=\linewidth]{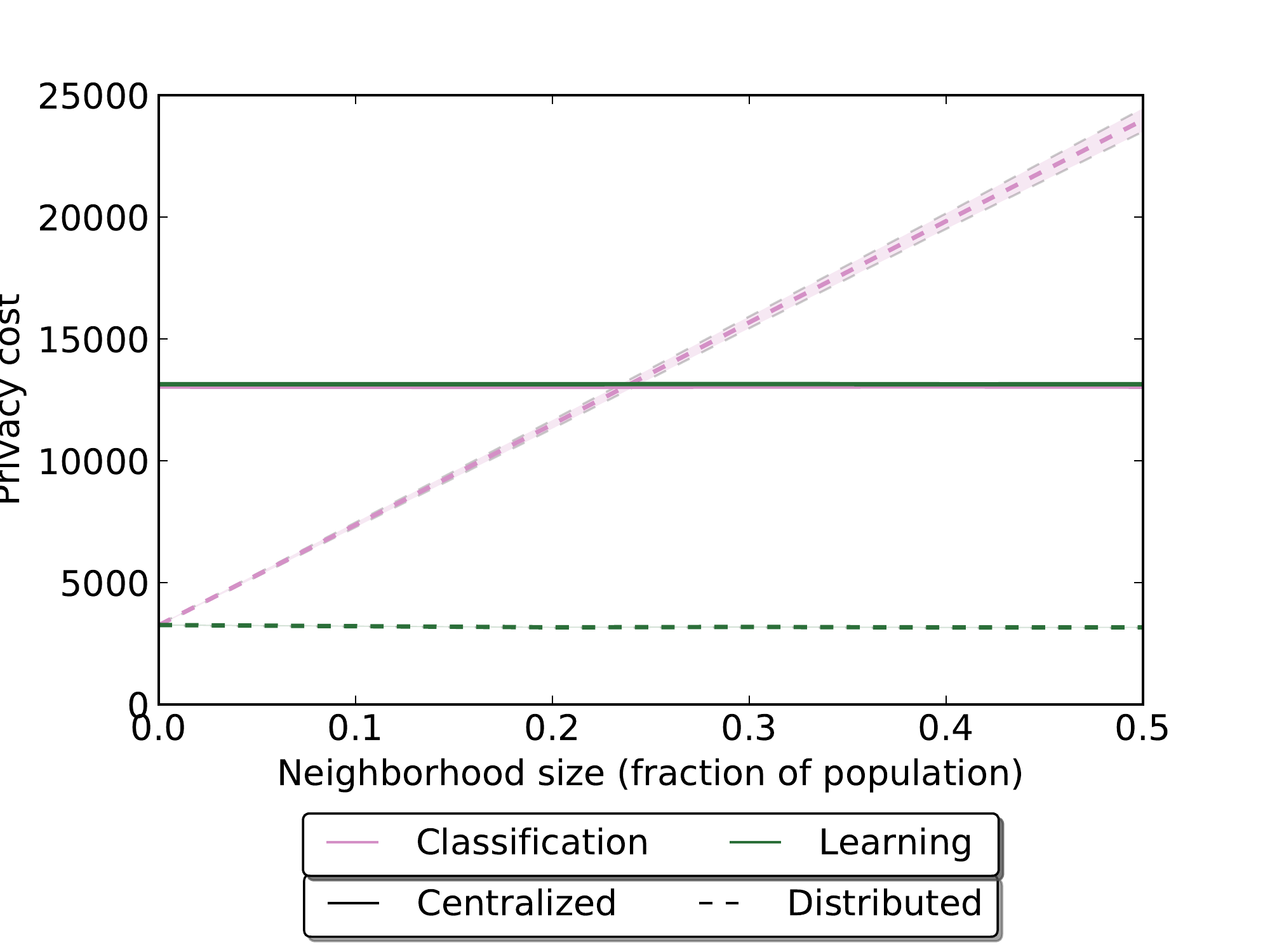}
  \end{minipage}
  \begin{minipage}[t]{0.45\linewidth}
\includegraphics[width=\linewidth]{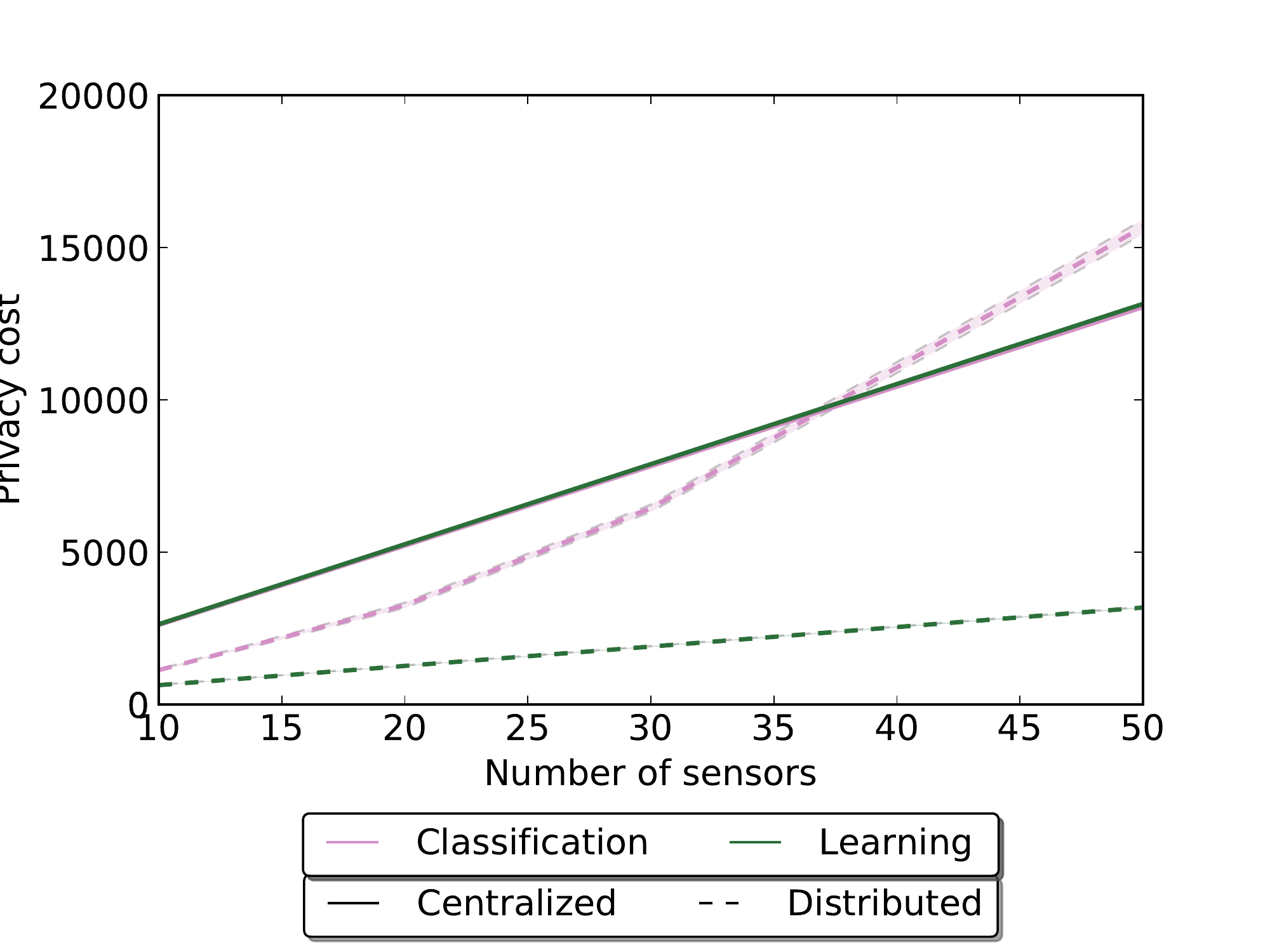}
  \end{minipage}
  \caption{Second experiment.  Comparison of privacy cost for different parameter configurations. Privacy of a centralized system does not vary when enabling learning. Shaded area corresponds to 0.95 confidence interval.}
  \label{fig:param3}
\end{figure}

\begin{figure}[t!]
  \centering
  \begin{minipage}[t]{0.45\linewidth}
\includegraphics[width=\linewidth]{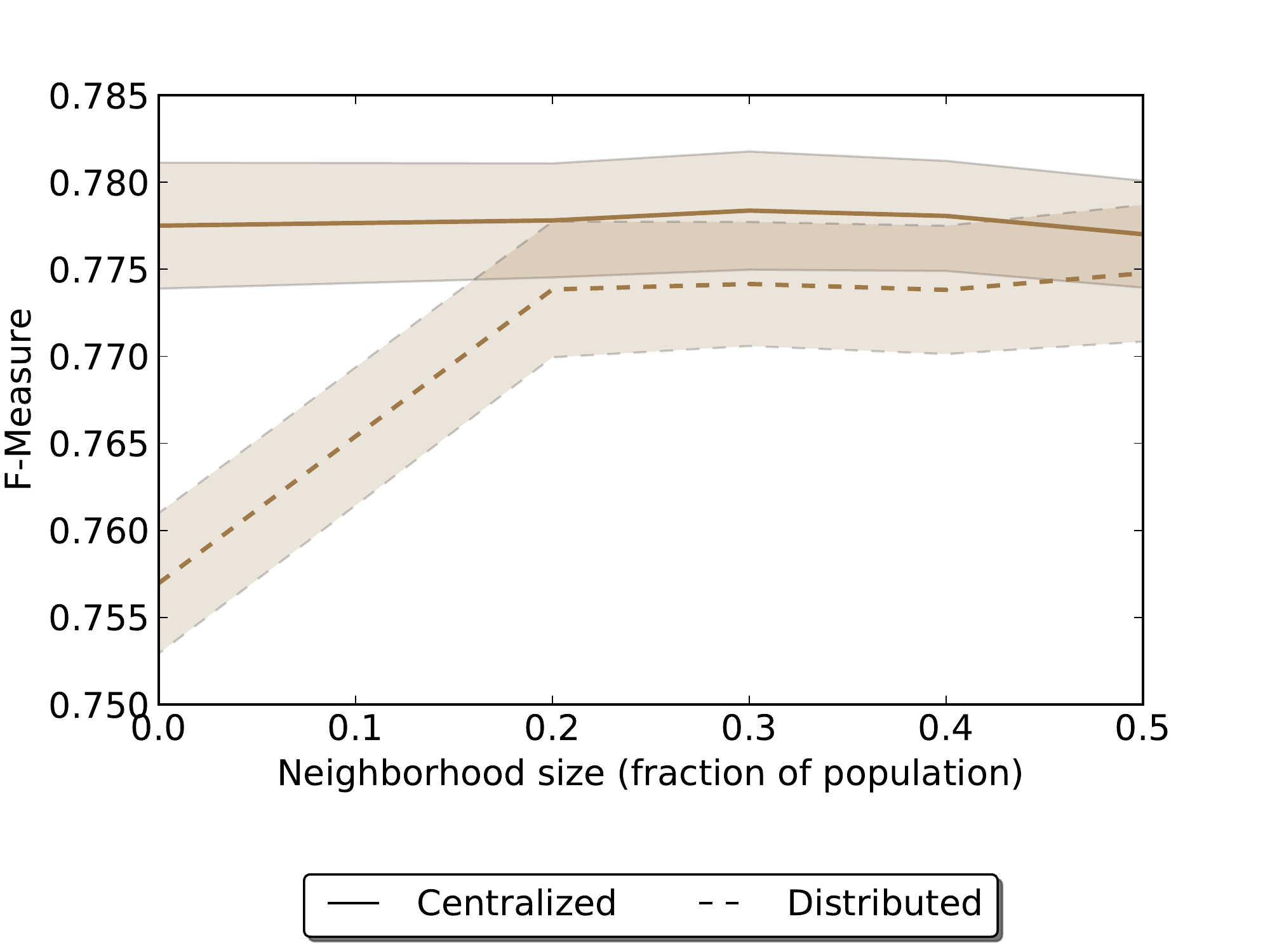}
  \end{minipage}
  \begin{minipage}[t]{0.45\linewidth}
\includegraphics[width=\linewidth]{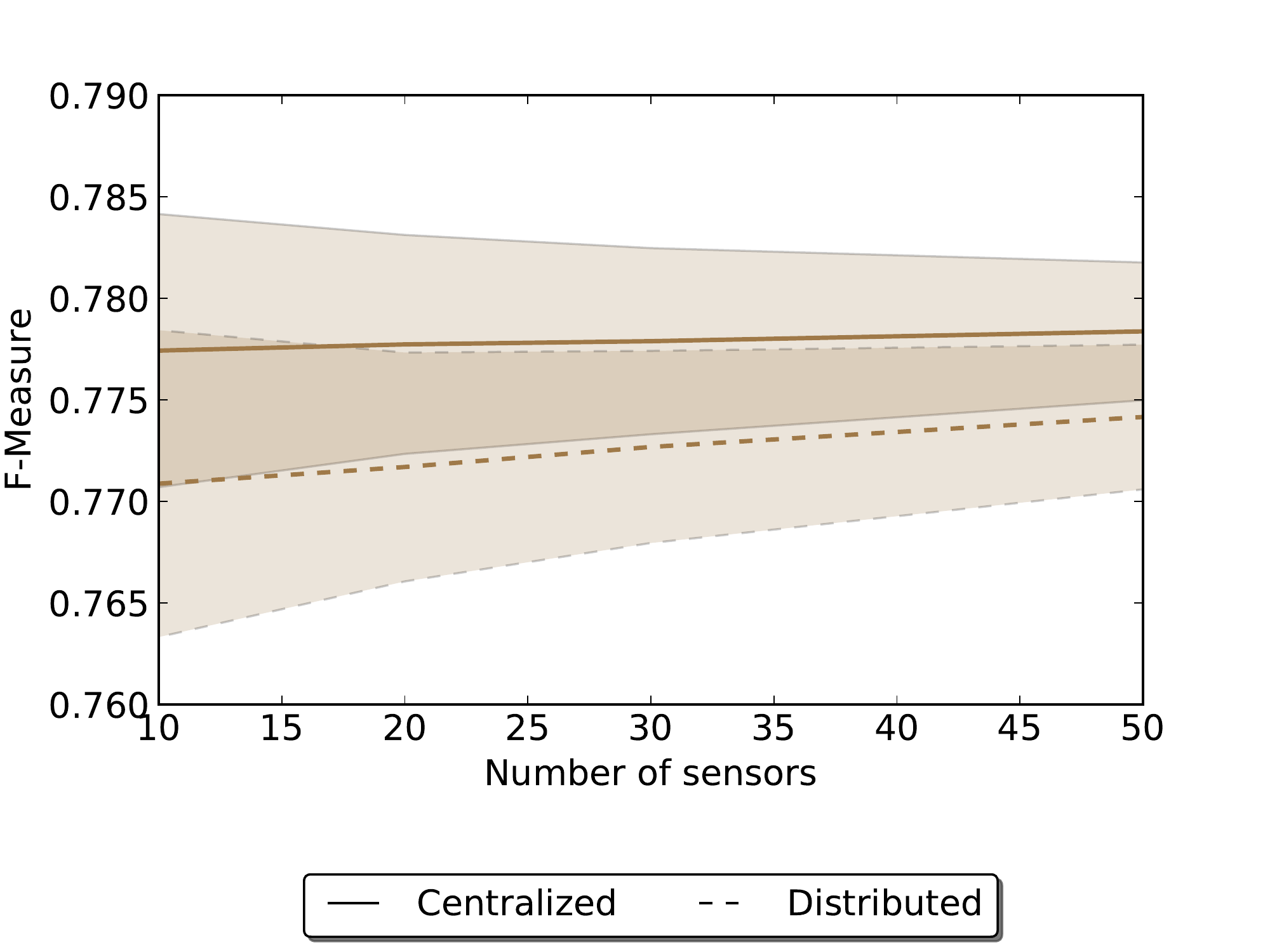}
  \end{minipage}
  \caption{Second experiment. Comparison of performance for different parameter configurations. Shaded area corresponds to 0.95 confidence interval.}
  \label{fig:param4}
\end{figure}

The \textbf{second experiment} looks at the effect of different parameter values on privacy and accuracy.
Figure \ref{fig:param1} shows the change in communication frequency towards the supervisor, for varying population and neighborhood size.
The neighborhood size is found not to change the communication frequency: an increase of the number of messages exchanged by the agents does not change the number of alarms sent to the supervisor.
The population size has instead a significant influence on the communication frequency, as each agent reports positive classifications to the supervisor, thus the more the agents the more the reports.
Note that a centralized and a fully distributed organization (cf. Figure \ref{fig:topo_cen} and \ref{fig:topo_distr}) have a similar communication footprint.
Figure \ref{fig:param2} shows the change in communication frequency towards the neighbors, in a distributed organization, for varying population and neighborhood size.
The plots highlight a positive dependency between communication frequency and both parameters, which is easily explained by the increased interconnectivity between the population: each agent communicates with its neighbors, so increasing either the number of agents or the number of neighbors increases the number of transmitted messages.
Figure \ref{fig:param3} shows that the privacy of a distributed organization (solid line) is lower than the privacy of a centralized organization (dashed line) whenever the neighborhoods are below a certain size. This result shows the tradeoff between centralized and distributed organizations: in a centralized organization the privacy cost is constant and comes from every sensor sending all its measurements to the central agent. Note that all this communication is not optimized by the learning algorithm, as it can only operate on the less numerous alarms that the agent reports, on which the learning algorithm can operate.
In a distributed organization the messages exchanged by sensors and agents do not contribute to the privacy cost, but so do messages exchanged between agents.
A distributed organization is less privacy-costly than a centralized organization whenever the messages transmitted between agents are less than the number of sensors.
Results show that a learning algorithm makes a distributed organization more privacy-preserving than a centralized organization, for any parameter values (cf. Figure \ref{fig:param3}).
To complete the evaluation, performance of a centralized and a fully distributed organization are compared, with learning enabled.
Figure \ref{fig:param4} shows that a distributed organization has slightly lower performance than a centralized organization, although the difference is not significant.
The only exception are \emph{decentralized} organizations (neighborhood size equals to zero) that have worse performance than a centralized organization.

\begin{figure}[t!]
  \centering
  \begin{minipage}[t]{0.45\linewidth}
\includegraphics[width=\linewidth]{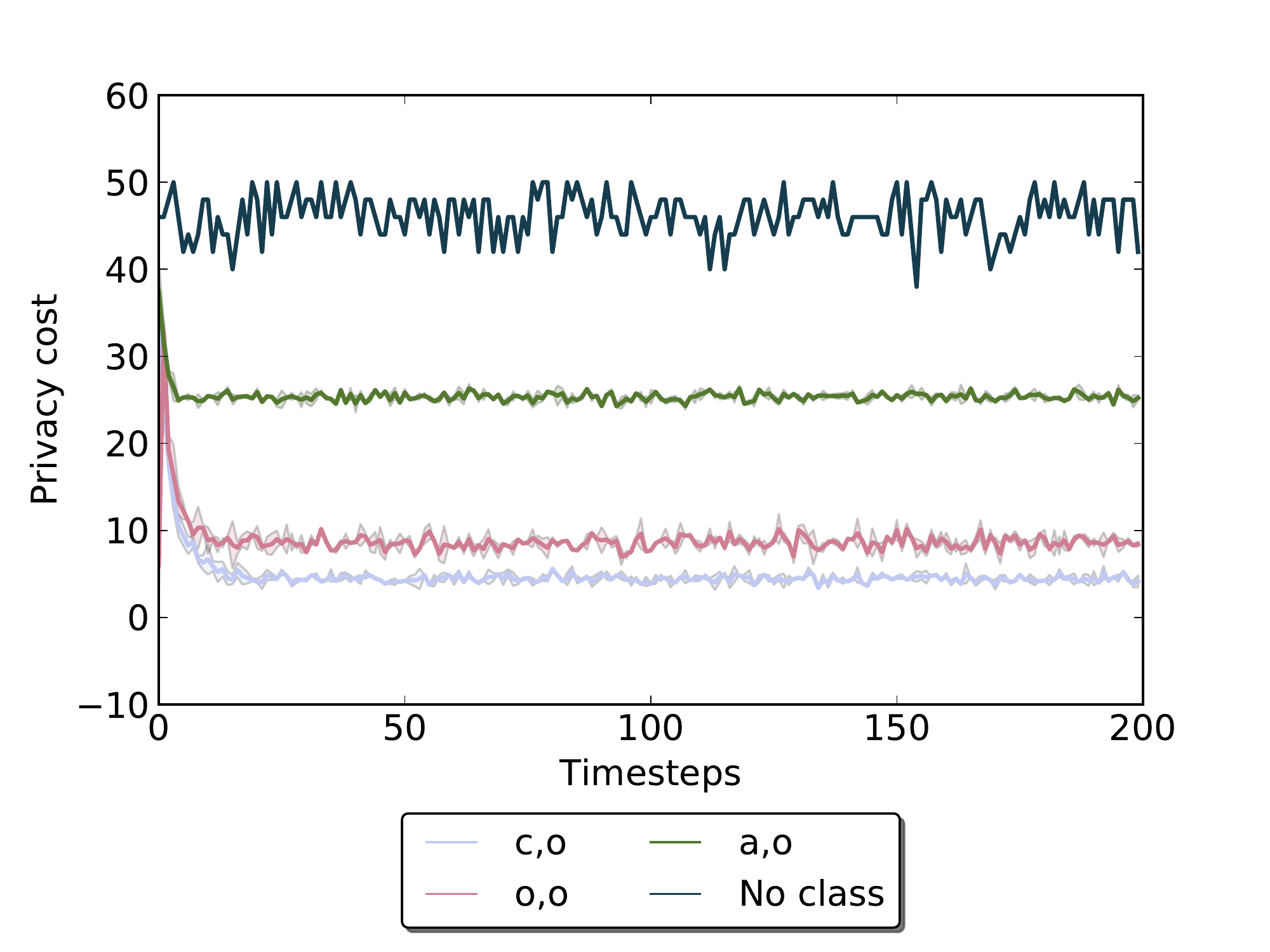}
\subcaption{Privacy cost}
  \end{minipage}
  \begin{minipage}[t]{0.45\linewidth}
    \includegraphics[width=\linewidth]{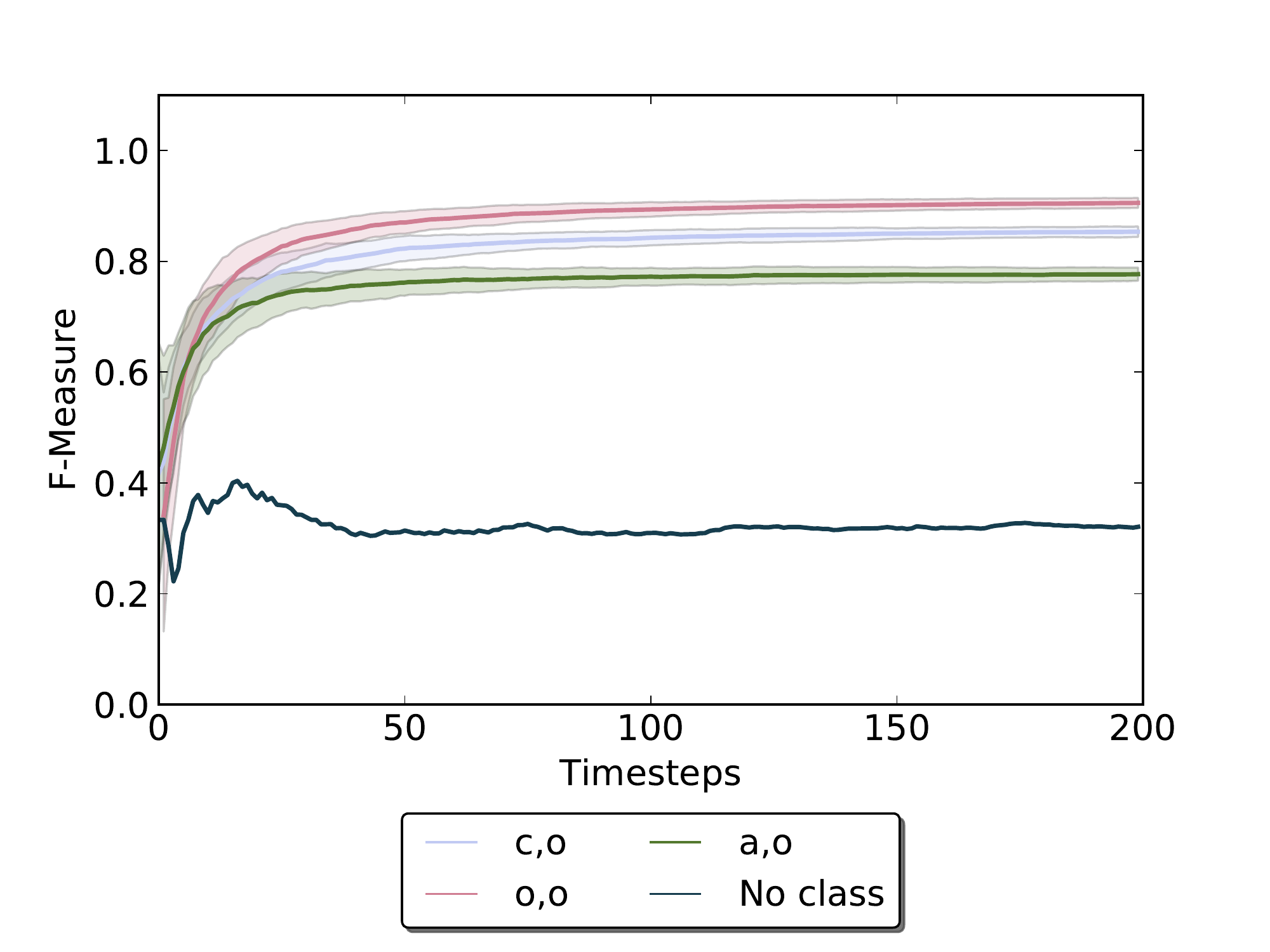}
\subcaption{Performance}
  \end{minipage}
  \caption{Third experiment. Privacy/performance trade-off of different communication functions. Shaded area corresponds to 0.95 confidence interval.}
  \label{fig:fct1}
\end{figure}

In the \textbf{third experiment} three communication criteria are evaluated: \emph{a} denotes transmitting all measurements, \emph{o} denotes transmitting only measurements with a positive classification, i.e. outliers, and \emph{c} denotes transmitting only measurements whose classification is uncertain i.e. confidence of classification is lower than a threshold.
Different criteria can be used to transmit to the neighbors and to the supervisor, a pair \emph{(x,y)} denotes using criterion x to communicate to the neighbors and criterion y to communicate to the supervisor. The baseline is denoted as ``No class'' and corresponds to a system which transmits all measurements.
Transmitting a message based on its classification increases the accuracy of the system and reduces the communication cost when compared to a system that transmits all messages.
Figure \ref{fig:fct1} shows a tradeoff between privacy and accuracy:
using the confidence as criterion, denoted with \emph{(c,o)}, gives the lowest privacy cost, while the combination \emph{(o,o)}, which uses classification label as criterion, produces the highest performance.

\section{Related Work}
\label{sec:rel_work}

This section compares the communication and privacy footprint of several state-of-the-art event detection algorithms in distributed sensor networks (cf. Table \ref{tab:1}).
Algorithms are implemented and evaluated within the simulation framework.
The performance obtained in the simulation are not comparable with those achieved by the original work, as all algorithms are run with the same type of classifier (one-class SVM), which might differ from what is used in the original work.
Nevertheless communication is modeled accurately, so the results can be used to compare the cost and privacy cost across algorithms.
For some algorithms the number of messages transmitted depends on the classification accuracy, thus having the same classifier across simulations helps reducing this variability by having a more uniform classification accuracy.
These algorithms are not chosen based on their accuracy but on their communication footprint.
Given that the interest is in the communication cost, more than in the performance of the algorithm, expanding the comparison to more algorithms with the same communication footprint would not strengthen the results.

\subsection{Communication Profiles}

The algorithms can be classified based on the criteria for communication to the supervisor and to the neighbors.
An algorithm is denoted as ``privacy preserving'' if the communication between neighbors does not reveal any privacy-sensitive information that could be intercepted by the supervisor, e.g. it is omitted or obfuscated.
All simulations are repeated for different values of parameters: number of sensors, probability of an event happening, and size of neighborhood (only for distributed organizations).

\begin{center}
\begin{tabular}{p{1.9cm}p{2.8cm}p{2.8cm}l}
privacy preserving& communication between agents & communication to supervisor & example\\
\hline
 True & outlier & outlier & \cite{zhang09_adapt_onlin_one_class_suppor}\\
 True & outlier & always & \\
 True & always & outlier & \cite{ruan08_binar}\\
 True & always & always & \\
\hline
 False & outlier & outlier & \cite{zhang12_statis_based_outlier_detec_wirel_sensor_networ}\\
 False & outlier & always & \\
 False & always & outlier & \cite{marin-perianu07_d_fler_distr_fuzzy_logic,wittenburg10}\\
 False & always & always & \cite{bahrepour10_distr_event_detec_wirel_sensor}\\
\hline
False & N/A & outlier & \cite{zoumboulakis07_escal,faulkner11,faulkner13}\\
False & N/A & always & \cite{bahrepour09_sensor_fusion_event_detec_wirel_sensor_networ}\\
\end{tabular}
\end{center}

Note that the communication frequency towards the supervisor varies with the number of agents but it does not vary with the size of the neighborhood, while the communication frequency towards the neighbors varies with both parameters. If agents have no neighbors, different communication methods between agents have no effect.

\subsection{Comparison}
\begin{figure}[H]
  \centering
  \begin{minipage}[t]{0.45\linewidth}
    \includegraphics[width=\linewidth]{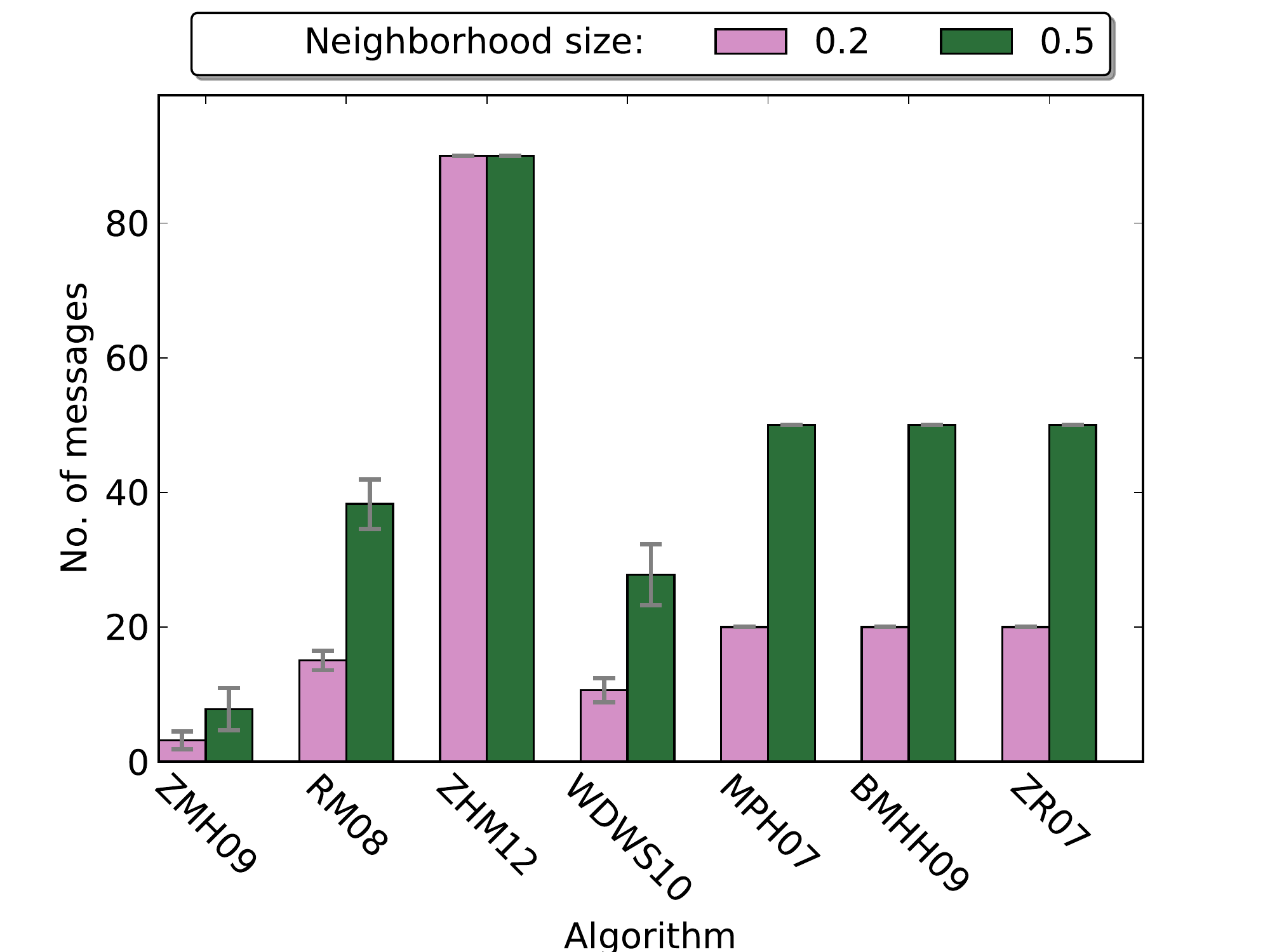}
  \end{minipage}
  \begin{minipage}[t]{0.45\linewidth}
\includegraphics[width=\linewidth]{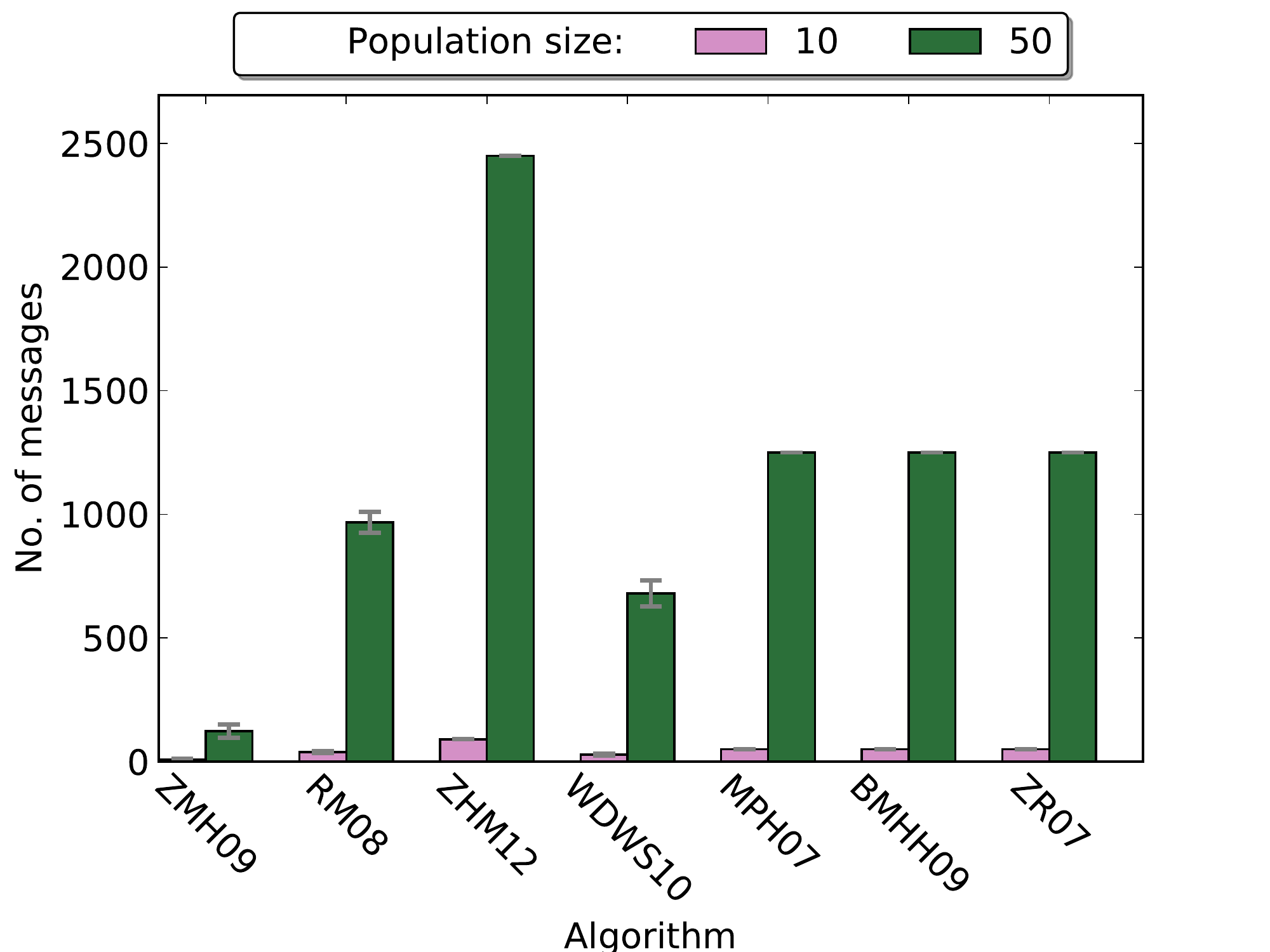}
  \end{minipage}
  \caption{Communication frequency towards the neighbors. Error bars correspond to 0.95 confidence interval.}
  \label{fig:lit1}
\end{figure}

\begin{figure}[H]
  \centering
  \begin{minipage}[t]{0.45\linewidth}
\includegraphics[width=\linewidth]{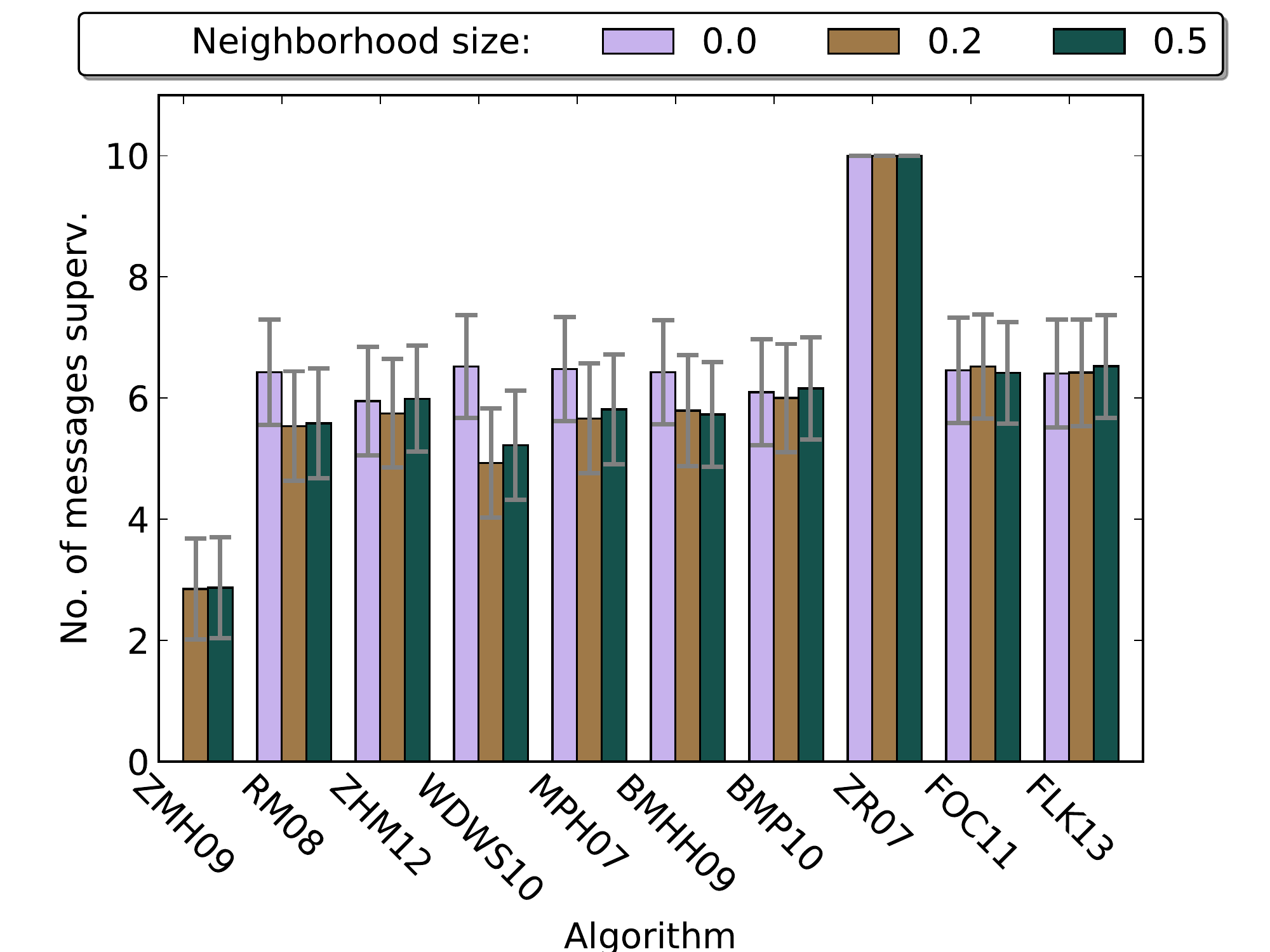}
  \end{minipage}
  \begin{minipage}[t]{0.45\linewidth}
\includegraphics[width=\linewidth]{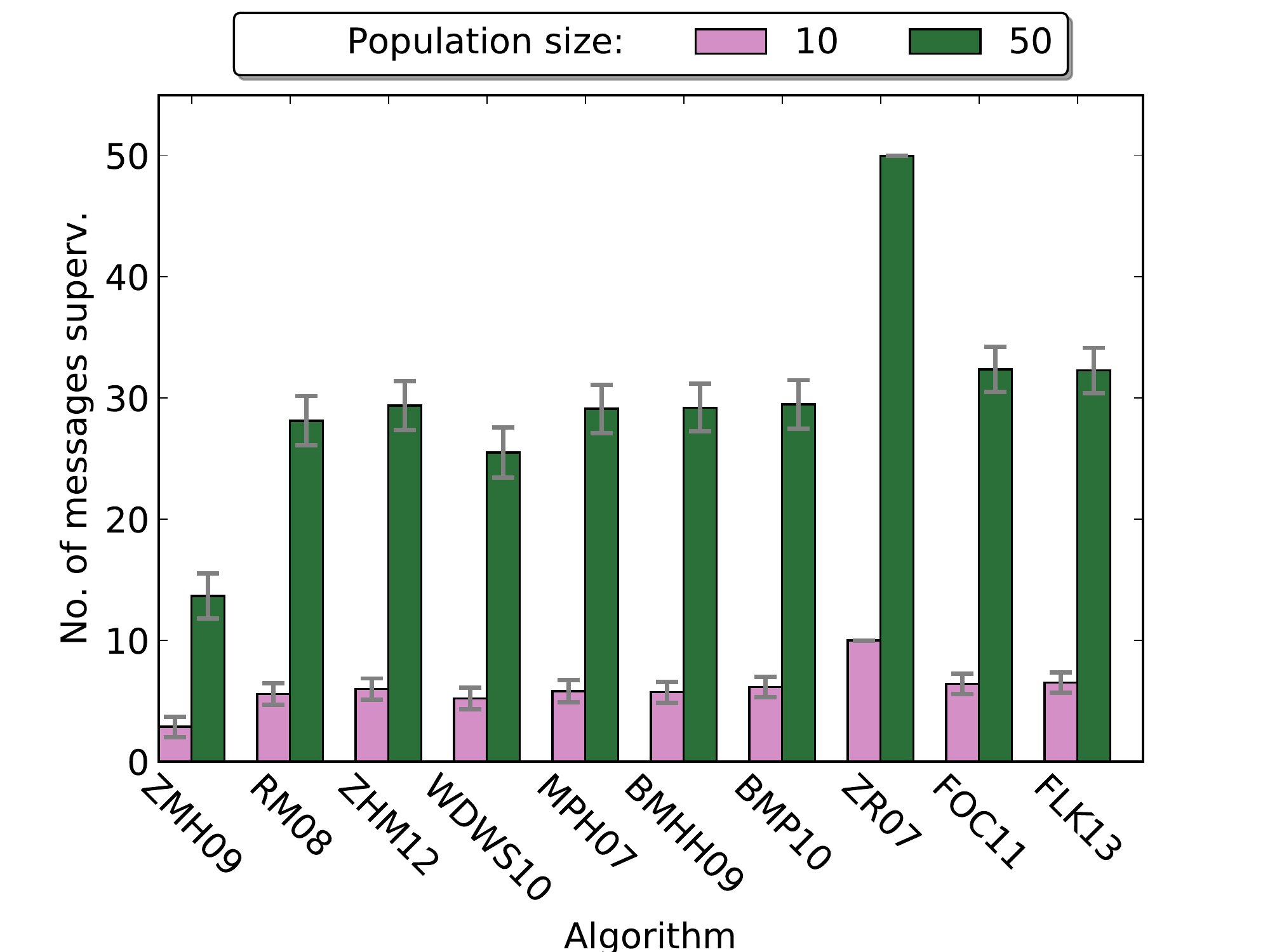}
  \end{minipage}
  \caption{Communication frequency towards the supervisor. Error bars correspond to 0.95 confidence interval.}
  \label{fig:lit2}
\end{figure}

\begin{figure}[H]
  \centering \includegraphics[width=0.5\linewidth]{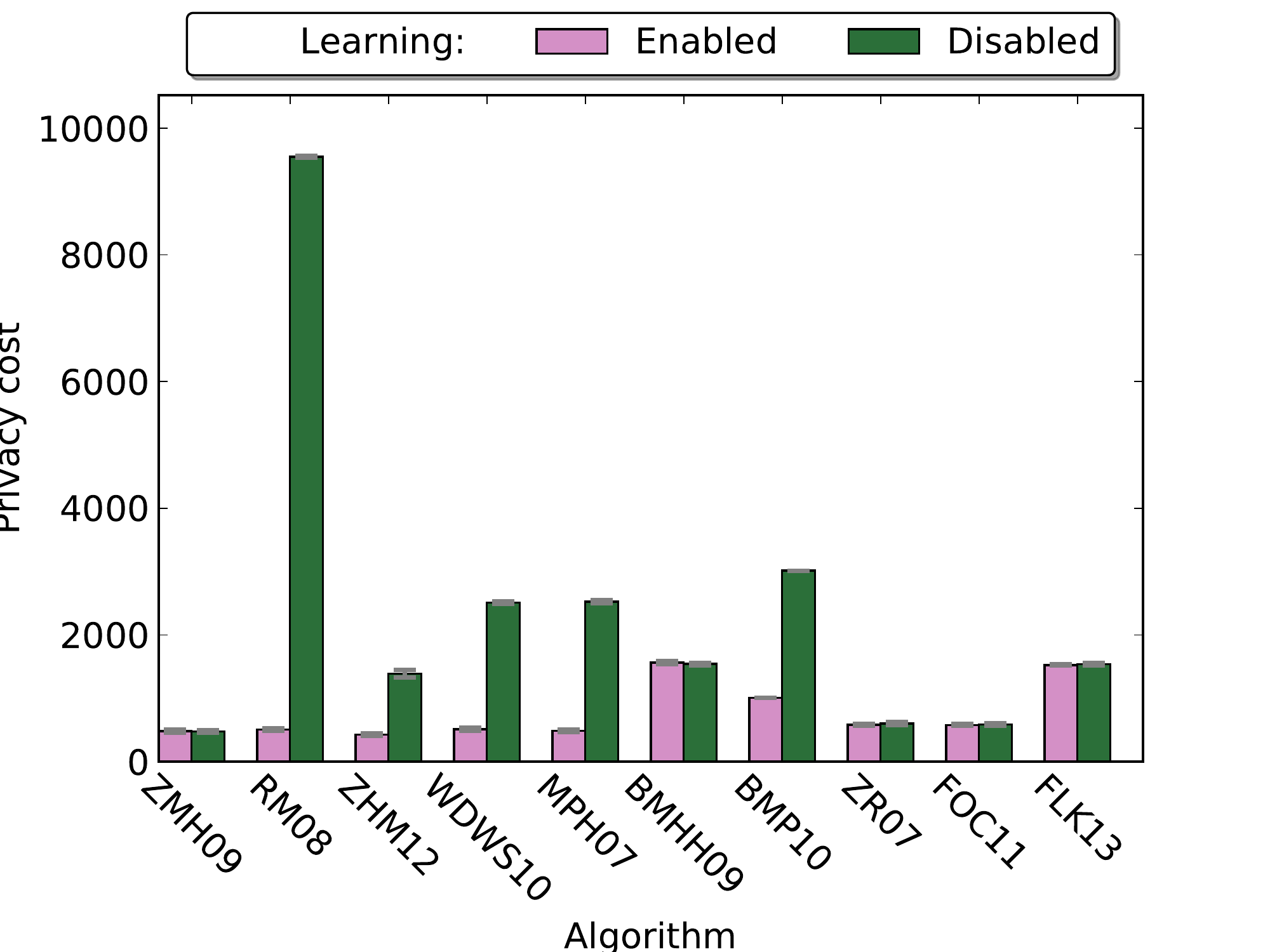}
  \caption{Effect of the learning mechanism on privacy cost. Simulation parameters: population size 10, neighborhood size 0.2. Error bars correspond to 0.95 confidence interval.}
  \label{fig:lit_learn}
\end{figure}

Figure \ref{fig:lit1} and Figure \ref{fig:lit2} show the average number of messages exchanged across simulations, error bars represent standard deviations.
Each point represents the value at the last iteration of a simulation. This representation has been chosen because values do not vary significantly during the simulation, so the final value is representative of the whole simulation.
The comparison shows that the frequency of communication varies broadly across algorithms, this result suggests that these algorithms were not designed with the goal of privacy minimization in mind.
Finally, the privacy cost produced by each of these algorithms is compared to the same algorithm equipped with PriMaL.
Figure \ref{fig:lit_learn} shows that PriMaL is able to reduce the privacy cost of the algorithms. Notable exceptions are algorithms that do not communicate to the neighbors and \emph{ZMH09}, which does so by transmitting privacy-neutral classification parameters instead of measurements.

\section{Conclusions}
\label{sec:discussion}
This paper introduces PriMaL, a machine-learning based mechanism to increase privacy in distributed sensor networks.
The mechanism can be combined with existing event detection algorithms and works on any network topology or organization.
The trade-off between its performance, communication frequency and privacy cost is quantified in a simulation framework.
The performance of PriMaL is validated by applying it on top of existing event detection algorithms and comparing their communication footprint and privacy cost.
Criteria for a generally-applicable event detection algorithm are individuated from this comparison.
The privacy-accuracy trade-off of the algorithm, equipped with PriMaL, is quantified for different parameter configurations and in decentralized, distributed and centralized network organizations.
An initial calibration phase might be required to train the learning algorithms.
Experimental results show that calibration has a temporary negative effect on detection performance, but this is compensated by an immediate reduction of the privacy cost.
Depending on the size of the neighborhoods, a distributed organization can have a lower communication footprint than a centralized organization.
It is found that the threshold under which a distributed organization has a lower privacy than the respective centralized organization is relatively low, thus in privacy-sensitive scenarios a centralized organization is preferable over a distributed organization with large neighborhoods.
Nevertheless, it is found that PriMaL is able to improve privacy of a distributed organization by optimizing the content of communication between agents.
PriMaL reduces the privacy cost of a distributed event detection algorithm below that of a centralized algorithm, given some assumptions about the protocol, while the performance of detection remains comparable to that of the centralized algorithm.

Future work should generalize the results by relaxing some assumptions, for example introducing correlation between measurements by allowing multiple sensors to perceive the same events.
The assumption of reliability should be relaxed and the system should be tested in the presence of failures of sensors and links.
Furthermore, the robustness the learning algorithm should be tested when adding uncalibrated nodes to the network.
The performance of the mechanism should also be investigated for more complex network topologies and more complex privacy settings.
For example where each agent associates individual privacy costs to transmittable information.
In this last scenario issues of cooperation, competition and negotiation might arise and should be investigated.


\section*{Acknowledgment}

The authors acknowledge support by the European Commission through the ERC Advanced Investigator Grant 'Momentum' [Grant No. 324247].

\section*{References}
\label{sec-14}
\newcommand{\etalchar}[1]{$^{#1}$}

\pagebreak
\appendix
\section{Experimental setup}
\label{sec:setup}

All results presented in this paper are an average of 50 simulations, which have been stopped after 200 iterations, enough to approach a steady state.

Performance of the system has been measured for the following parameters values:
\begin{itemize}
\item number of sensors: 10 and 50
\item size of neighborhood: 0\%, 20\% and 50\% of the population size.
\item frequency of events:
\begin{itemize}
\item probability of an event happening at every sensor/timestep: 0.2 and 0.5
\item total number of events happening during the simulation: 300 and 600
\end{itemize}
\end{itemize}

Results show that the effect of the frequency of events is straightforward: the more frequent the events, the higher the communication and the higher the precision of the classification.

All graphs compare simulations with the same parameter configuration.

The framework is implemented in Python, it is composed by the following entities:

\subsection{Sensors}
A sensor reads and transmits measurements. Each sensor has a type, identifying the events it can measure.

At each timestep a sensor measures the environment.
The measurement is sampled from one of two random variables, one for normal and one for abnormal measurements. Depending on the desired type of measurement, one or the other function is used.

Each sensor contains a list of agents that are subscribed to it, to whom it it transmits every measurement.
\subsection{Network}
Sensors and agents transmit their messages through the network.
The task of the network is to compute the costs of each transmission, which depend on the message sent and on the distance between source and destination.
The network contains a topology matrix that identifies the distance between each sensor and agent. The distance is binary: either 0 (pair is physically connected) or 1 (pair is connected through the network). This topology determines the cost of transmission between two entities.
\subsection{Remote}
This object represents the supervisor. Its task is to provide the ground truth about events, record the alarms and compute the system accuracy.
\subsection{Transmitter}
Transmitter uses Q-Learning to learn what information to transmit. Each agent holds two instances of transmitter: one for transmissions to other agents, the other for transmissions to the supervisor. Transmitter receives as feedback the costs computed by the network, and learns what information is good to transmit and what information can be skipped.
\subsection{Classifier}
It encapsulates a classification algorithm (SVM) and is used by an agent to classify its measurements. Each agent trains a classifier with the local measurements and uses its classification to decide when to send alarms.
The classifier provides the classification confidence as well, depending on which an agent might communicate with its neighbor to ask for their opinion on the event. It can work is an unsupervised or in a supervised way, depending on whether the feedback is provided.
\subsection{Agent}
The main entity of the simulation. The agent receives and processes measurements from a number of sensors.
Each agent contains a classifier for each sensor type it is subscribed to. Each classifier is trained on the measurements reported by all sensors of that type that the agent is subscribed to.
Agents contain a list of neighbors to which they can communicate. Communication consists in sending a measurement to the neighbors and receiving their classification for that value, which are then aggregated into a single classification label.
Communication with the neighbors and with the supervisor is performed whenever some conditions are met, by default whenever the measurement is classified as event.

Agents are equipped with a single classifier for each sensor type, which is trained in a supervised way on the ground truth provided by the supervisor.
Each agent contains two transmitters, one for communications to the neighbors, the other for communications to the supervisor. The transmitters learn independently what pieces of information to transmit and what to suppress.

\subsection{Classification}

Classification is implemented using the One-class SVM class in the SKlearn library. One-Class SVM has been chosen because it is the state of the art classification, and they can be used both in supervised and unsupervised mode, giving us implementation flexibility. This choice does not influence the applicability of the mechanism.

Unless events are extremely rare, SVMs cannot be successfully trained online without supervision: the first measurements that the classifier receives have a strong influence on the classification, so if an event is measured in the early timesteps the classification accuracy might be compromised. A few labeled elements (normal measurements) are enough to initialize the classifier correctly.

Training should be done including only one class of elements, either normals or outliers. An SVM is able to learn online and tolerate some false measurements, assuming the initialization was correct. The classifier is trained with normal points.

\subsection{Learning}

In the experiments the cost function is arbitrary, where only one piece of information has a privacy cost. This is realistic for the scenario of earthquake detection with smartphones, where only the location is privacy-sensitive.
All other pieces of information have a unitary transmission cost.

The choice of cost function influences the costs and learning, but it does not influence the applicability of the mechanism.

The learning algorithm is trained using a combination of a fixed transmission cost and a variable privacy cost that can differ based on the content of the messages.

Q-Learning has been chosen because it is a online supervised learning algorithm that is computationally very fast for small spaces.
Moreover the algorithm has not been optimized, to prove the point that even a simple and not efficient solution can bring a benefit to the system, which can become even greater for more sophisticated learning algorithms. This choice does not influence the applicability of the mechanism.

\subsection{implementation assumptions}
\begin{itemize}
\item classification is binary, as in \cite{faulkner13,faulkner11,zhang09_adapt_onlin_one_class_suppor,ruan08_binar}, this assumption is not a necessary precondition for this mechanism to operate.
\item In the distributed setting agents are computing the overall classification of the neighborhood using majority vote, similarly to \cite{ruan08_binar,bahrepour09_use,bahrepour10_distr_event_detec_wirel_sensor}. This aggregation function has been chosen arbitrarily, but any other function will work with the mechanism.
\item Each sensor has its own classifier, as in  \cite{faulkner11,zhang09_adapt_onlin_one_class_suppor,ruan08_binar,bahrepour10_distr_event_detec_wirel_sensor}. Other approaches keep a system-level threshold \cite{faulkner13,bahrepour09_sensor_fusion_event_detec_wirel_sensor_networ}, which is not suitable in this setting as it reduces the flexibility of the system.
\begin{itemize}
\item Classification is implemented with One-Class SVM classifiers \cite{laskov04_intrus_detec_unlab_data_with,zhang09_adapt_onlin_one_class_suppor} with a sliding window, as they are robust and able to operate both in a supervised and unsupervised way. The choice of classifier is arbitrary.
\item The classifier is trained on the normal data and detects events as outliers, as in \cite{faulkner11}.
\item If a sensor is connected to more than one agent, each agent keeps a separate classifier.
\end{itemize}
\end{itemize}


\section{Methodological Requirements}
An algorithm must have the following properties in order to be generally applicable, i.e. work in every network organization:
\begin{itemize}
\item Sensors specifications (inputs and outputs) cannot be modified.
\item Sensors have no intelligence
\item Each agent can be associated with one or more sensors, this allows the method to be independent of the network organization
\item Have independent classifiers in each agent
\item Work without offline training
\item Agents are allowed to communicate with neighbors, for example if classification confidence is low.
\end{itemize}

\section{Assumptions}
For simplicity of implementation, the following is assumed:
\begin{itemize}
\item Events are independent. With this assumption heterogeneous networks can be decomposed in independent networks for each type of event and sensor type, which can be analyzed individually.
  \begin{itemize}
  \item Time correlation is not considered, but it is neither excluded.
  \item There is no correlation between signals of different sensor types, e.g. fire cannot be detected by both temperature and smoke sensors.
  \item One sensor type cannot detect more than one measurement type, e.g. temperature sensors cannot detect both fire and intrusion, this can be modeled by one virtual sensor type for each event type.
  \end{itemize}
\item Each sensor is connected to only an agent. A sensor connected to multiple agents can be modeled as several virtual sensors with the same readings.
  \begin{itemize}
  \item The sensors are perfectly reliable and always communicate the truth.
  \end{itemize}
\item Agents can communicate with others and ask for their opinion.
  \begin{itemize}
  \item The network is perfectly reliable: transmissions are always successful.
  \item Agents do not forward messages from other agents, this avoids loops and cascade effects.
  \end{itemize}
\item Everything that is transmitted over the network is subject to a (privacy and communication) cost.
\item The location of a measurement cannot be inferred by knowing what an agent transmits, even if that agent is connected only to only one sensor. This assumption is reasonable if agents randomly change their ID at every transmission e.g. in mix-nets.
\end{itemize}

\clearpage
\section{Detailed review of event detection algorithms}

\begin{wraptable}{r}{0.4\textwidth}
    \scriptsize
  \sbox0{
    \begin{tabular}{|l|l|l|p{1.8cm}|l|p{1.8cm}|p{0.8cm}|p{0.8cm}|}
\\
Paper & classification & organization & reduce comm. to neighbors & supervisor & reduce comm. to supervisor & private comm. & trained\\
\hline
Proposed & by agent & distributed & variable & logging/feedback & variable & yes & online\\
\cite{zhang09_adapt_onlin_one_class_suppor} & by agent & distributed & only outliers & logging & only outliers & yes & online\\
\cite{ruan08_binar} & group consensus & distributed & no & logging & only outliers & yes \footnotemark{} & online\\
\cite{zhang12_statis_based_outlier_detec_wirel_sensor_networ} & by agent & distributed & only outliers & logging & only outliers & no & offline\\
\cite{wittenburg10} & by agent & distributed & no & logging & only outliers & no & offline\\
\cite{marin-perianu07_d_fler_distr_fuzzy_logic} & by agent & distributed & no & logging & only outliers & no & offline\\
\cite{bahrepour10_distr_event_detec_wirel_sensor} & central consensus & decentralized & no & consensus & no & no & offline\\
\cite{bahrepour09_sensor_fusion_event_detec_wirel_sensor_networ} & central consensus & hierarchical & N/A & consensus & no & no & offline\\
\cite{zoumboulakis07_escal} & by sensor & decentralized & N/A & logging & only outliers & no & offline\\
\cite{faulkner11} & by sensor & decentralized & N/A & hypothesis testing & only outliers & no & offline\\
  \cite{faulkner13} & at central level & centralized & N/A & event detection & only outliers \footnotemark{} & no \footnotemark{} & offline\\
    \end{tabular}
  }
  \rotatebox{90}{%
    \parbox{0.45\textheight}{
          \centering
  \caption{Comparison of outlier-detection approaches}
  \label{tab:1}
\endgraf\bigskip
\usebox0}
}
\normalsize
\end{wraptable}
 \addtocounter{footnote}{-3} 
 \stepcounter{footnote}\footnotetext{although every sensor measures the same event}
 \stepcounter{footnote}\footnotetext{if using hierarchical anomaly detection}
 \stepcounter{footnote}\footnotetext{unless using in-network aggregation}


\noindent
\begin{minipage}{\linewidth}
\cite{faulkner11} develop a system for detecting events from smartphones, which imposes some extra constraints on communication and robustness to noise.
Their default system is decentralized, but not distributed: agents do not exchange any information with their neighbors.
Nevertheless they introduce a way of making the algorithm distributed by having the network compute the detection in a bottom-up way, by propagating the signal up the hierarchy. This works well if there is locality because the nodes close in the hierarchy will also be close in the environment, if locality is absent the communication overhead is more intensive.
Each sensor has its own classifier, trained on the local observations and triggers binary alarms anytime a measurement is classified as event.
The accuracy of their method is based on the assumption that event are very rare, so each sensor is able to identify events based purely on its own local observations without requiring any supervision.

\cite{faulkner13} looks at event detection in noisy sensor networks, specifically at earthquake detection.
Their sensors generate noisy binary measurements: classification of measurements (as event or non-event) is flipped with some constant probability.
They introduce a centralized fusion center to aggregate individual measurements. The fusion center learns a global threshold which is applied to all sensors, therefore individual sensors cannot adapt the threshold to the local setting.
\end{minipage}
\clearpage
They also introduce a hierarchical aggregation mechanism which reduces the communication footprint by computing the statistics in a bottom-up way. This mechanism can be applied only if the learned basis exhibits a hierarchical structure.
The amount of information transmitted to the fusion center varies based on the system configuration: with the default setting all sensors send all their measurements to the fusion center, while when implementing ``hierarchical anomaly detection'', only what is classified as event by the local classification algorithm is transmitted.

\cite{zhang09_adapt_onlin_one_class_suppor} develop a distributed system where each sensor is equipped with a One-class SVM classifier that is trained online on the local observations.
They assume each sensor is equipped with a classifier. By not separating the roles of sensors and agents, they effectively limit the flexibility of the organization as the system cannot assume any configuration in which an agent supervises more than one sensor, e.g. centralized, hierarchical.
They assume that measurements are spatially correlated, therefore neighbors have similar classifications. In this case spatial correlation is a feature of the environment, i.e. sensors that are geographically close together measure similar events, but the same situation could be obtained with an appropriate network topology, i.e. sensors that are topologically close measure similar events.
Sensors exploit the correlation in the data by exchanging parameters with their neighbors, which they use to classify the events they measure. Sensors broadcast their updated parameter anytime they detect an outlier, each agent aggregates the parameters sent by the neighbors and use them to confirm their classifications.
Moreover in this approach classifiers are isolated, in the sense that the parameters exchanged by the neighbors are used after the classification as a further check.
With this system calibration is not possible: a classifier with a bad accuracy will not be able to improve by learning from the neighbors.

\cite{zhang12_statis_based_outlier_detec_wirel_sensor_networ} uses statistical models, which are trained offline, to detect outliers.
Subsequent sensor readings are collected in time series and outliers are defined as a sequence of observations that fall outside the predicted confidence interval.
With respect to the other solutions, the definition of outlier changes from defining a single anomalous observation to defining a sequence of observations.
Every agent asks confirmation to the neighbors whenever it detects an outlier, and uses this information to verify its classification.

\cite{ruan08_binar} look at the consensus problem in agreeing on a global classification label.
This system differs from others as agents have to take not one but several decisions, considering the opinion of the neighbors, until the majority of agents agrees on an answer.
This system is very demanding in terms of transmission cost, as each sensor communicates its classification to the whole population (fully connected graph) at every iteration, until (and if) consensus is reached.
An assumption behind this system is that all sensor perceive the same events and collectively classify them, while other distributed systems only require that some other agent perceives events of the same type.

\cite{bahrepour09_sensor_fusion_event_detec_wirel_sensor_networ} also look at the consensus problem. The difference with the previous approach is the presence of a fusion center, which collects classifications from agents and computes consensus.
It assumes the event (fire) has a known signature that can be learned offline, so the system effectively does not need to adapt to changes.
Sensors send their measurements to sensor nodes, what we call agents, which compute classification and report only the outliers to the fusion center.
The fusion center distinguishes between outliers occurring at individual nodes and events that are detected by many nodes, this is based on the assumption that events are spatially and temporally correlated, so an event can be detected by more than one node.
Agents classify independently and do not communicate with each other, so this is a typical example of hierarchical configuration.

\cite{bahrepour10_distr_event_detec_wirel_sensor} is an extension of the previous work with the inclusion of a reputation mechanism.
Each sensor type has a decision tree that is used to classify the events, this algorithm needs to be trained offline and the training set is the same over many sensors.
Such requirement does not allow sensors to specialize to their individual environmental conditions, for example background noise.
The reputation system, built on the evaluations given by agents to each other, is useful to detect and neutralize broken sensors, but has the disadvantage of increasing the communication overhead: the supervisor requires every agent to transmit the reputation scores they assign to every other neighbor, this sums up to the communication between neighbors.
Despite neighbors communicate with each other their measurements, the system is decentralized as this communication is only used to build the reputation scores and not to improve classification. Moreover there is a central aggregator which computes the global consensus.

\cite{wittenburg10} develop a distributed approach where nodes improve their classification with ``feature vectors'', i.e. a compressed version of the measurement, sent by the neighbors.
Communication overhead is kept low by reporting only events that are detected, although neighbors communicate at every timestep.
This approach requires an offline training phase where a centralized feature selection algorithm is used to simplify and summarize the measurements coming from the sensors.
This, and logging the reported events, are the only tasks the central entity performs.
At the individual level, every agent has its own classifier so it is able to adapt to local differences. An assumption is that events propagate evenly to all neighbors.

\cite{zoumboulakis07_escal} perform event detection with multiple sensors. They convert measurements using the Symbolic Aggregate Approximation (SAX) algorithm, and perform pattern matching on the symbols.
Their algorithm features an offline learning phase where a signature of the event is built by performing feature selection. Sensors classify in isolation, so their approach is decentralized but not distributed.

\cite{marin-perianu07_d_fler_distr_fuzzy_logic} develop a distributed fuzzy engine for event detection. Fuzzy logic is able to process multi-sensor data, so individual data is combined to that of the neighbors in order to improve the classification accuracy. A disadvantage of this system is that it is very communication intensive, as neighbors broadcast their measurements at every timestep.
In this approach reasoning is decoupled from the sensors, so in principle an agent can serve any number of sensors, which makes this algorithm adaptable to different network organizations.




\begin{thebibliography}{dMHVB13}

\bibitem[BMHH09]{bahrepour09_sensor_fusion_event_detec_wirel_sensor_networ}
Majid Bahrepour, Nirvana Meratnia, Paul J.~M. Havinga, and M.~Havinga.
\newblock Sensor fusion-based event detection in wireless sensor networks.
\newblock In {\em Proceedings of the 6th Annual International Conference on
  Mobile and Ubiquitous Systems: Computing, Networking and Services}, 2009.

\bibitem[BMP{\etalchar{+}}10]{bahrepour10_distr_event_detec_wirel_sensor}
Majid Bahrepour, Nirvana Meratnia, Mannes Poel, Zahra Taghikhaki, and Paul~J.M.
  Havinga.
\newblock Distributed event detection in wireless sensor networks for disaster
  management.
\newblock In {\em International Conference on Intelligent Networking and
  Collaborative Systems}, 2010.

\bibitem[BS04]{beresford04_mix}
A.R. Beresford and F.~Stajano.
\newblock Mix zones: user privacy in location-aware services.
\newblock In {\em Proceedings of the Second IEEE Annual Conference on Pervasive
  Computing and Communications}, 2004.

\bibitem[BZMH09]{bahrepour09_use}
Majid Bahrepour, Yang Zhang, Nirvana Meratnia, and Paul~J.M. Havinga.
\newblock Use of event detection approaches for outlier detection in wireless
  sensor networks.
\newblock In {\em International Conference on Intelligent Sensors, Sensor
  Networks and Information Processing (ISSNIP)}, 2009.

\bibitem[DK05]{duckham05_formal_model_obfus_negot_locat_privac}
Matt Duckham and Lars Kulik.
\newblock {\em A Formal Model of Obfuscation and Negotiation for Location
  Privacy}, pages 152--170.
\newblock Lecture Notes in Computer Science. Springer Science + Business Media,
  2005.

\bibitem[dMHVB13]{montjoye13_unique_crowd}
Yves-Alexandre de~Montjoye, C{\'e}sar~A. Hidalgo, Michel Verleysen, and
  Vincent~D. Blondel.
\newblock Unique in the crowd: The privacy bounds of human mobility.
\newblock {\em Scientific Reports}, 3, 2013.

\bibitem[FLK13]{faulkner13}
Matthew Faulkner, Annie~H. Liu, and Andreas Krause.
\newblock A fresh perspective.
\newblock In {\em Proceedings of the 12th international conference on
  Information processing in sensor networks - IPSN}, 2013.

\bibitem[FOC{\etalchar{+}}11]{faulkner11}
Matthew Faulkner, Michael Olson, Rishi Chandy, Jonathan Krause, K~Mani Chandy,
  and Andreas Krause.
\newblock The next big one: Detecting earthquakes and other rare events from
  community-based sensors.
\newblock In {\em 10th International Conference on Information Processing in
  Sensor Networks (IPSN)}, pages 13--24. IEEE, 2011.

\bibitem[Gol05]{goldreich05_found_crypt_primer}
Oded Goldreich.
\newblock Foundations of cryptography - a primer.
\newblock {\em Foundations and Trends in Theoretical Computer Science},
  1(1):1--116, 2005.

\bibitem[LDXZ15]{li2015internet}
Shancang Li, Li~Da~Xu, and Shanshan Zhao.
\newblock The internet of things: a survey.
\newblock {\em Information Systems Frontiers}, 17(2):243--259, 2015.

\bibitem[LSKM04]{laskov04_intrus_detec_unlab_data_with}
P.~Laskov, C.~Sch{\"a}fer, I.~Kotenko, and K.-R. M{\"u}ller.
\newblock Intrusion detection in unlabeled data with quarter-sphere support
  vector machines.
\newblock {\em PIK - Praxis der Informationsverarbeitung und Kommunikation},
  27(4):228--236, 2004.

\bibitem[MBG{\etalchar{+}}15]{minson15_crowd_earth_early_warnin}
S.~E. Minson, B.~A. Brooks, C.~L. Glennie, J.~R. Murray, J.~O. Langbein, S.~E.
  Owen, T.~H. Heaton, R.~A. Iannucci, and D.~L. Hauser.
\newblock Crowdsourced earthquake early warning.
\newblock {\em Science Advances}, 1(3), 2015.

\bibitem[MPH07]{marin-perianu07_d_fler_distr_fuzzy_logic}
Mihai Marin-Perianu and Paul Havinga.
\newblock {\em D-FLER - A Distributed Fuzzy Logic Engine for Rule-Based
  Wireless Sensor Networks}, pages 86--101.
\newblock Ubiquitous Computing Systems. Springer Science + Business Media,
  2007.

\bibitem[MYYR13]{ma13_privac_vulner_publis_anony_mobil_traces}
Chris Y.~T. Ma, David K.~Y. Yau, Nung~Kwan Yip, and Nageswara S.~V. Rao.
\newblock Privacy vulnerability of published anonymous mobility traces.
\newblock {\em IEEE/ACM Transactions on Networking}, 21(3):720--733, 2013.

\bibitem[PNV{\etalchar{+}}16]{pournaras16_self_regul_infor_sharin_partic_social_sensin}
Evangelos Pournaras, Jovan Nikolic, Pablo Vel{\'a}squez, Marcello Trovati, Nik
  Bessis, and Dirk Helbing.
\newblock Self-regulatory information sharing in participatory social sensing.
\newblock {\em EPJ Data Science}, 5(1):1, 2016.

\bibitem[RM08]{ruan08_binar}
Yongxiang Ruan and Yasamin Mostofi.
\newblock Binary consensus with soft information processing in cooperative
  networks.
\newblock In {\em 47th IEEE Conference on Decision and Control}, 2008.

\bibitem[RSCB13]{rawat13_wirel_sensor_networ}
Priyanka Rawat, Kamal~Deep Singh, Hakima Chaouchi, and Jean~Marie Bonnin.
\newblock Wireless sensor networks: a survey on recent developments and
  potential synergies.
\newblock {\em The Journal of Supercomputing}, 68(1):1--48, 2013.

\bibitem[SNQ12]{shahid12_charac_class_outlier_detec_techn}
Nauman Shahid, Ijaz~Haider Naqvi, and Saad~Bin Qaisar.
\newblock Characteristics and classification of outlier detection techniques
  for wireless sensor networks in harsh environments: a survey.
\newblock {\em Artificial Intelligence Review}, 43(2):193--228, 2012.

\bibitem[WADX15]{whitmore2015internet}
Andrew Whitmore, Anurag Agarwal, and Li~Da~Xu.
\newblock The internet of things—a survey of topics and trends.
\newblock {\em Information Systems Frontiers}, 17(2):261--274, 2015.

\bibitem[WDWS10]{wittenburg10}
Georg Wittenburg, Norman Dziengel, Christian Wartenburger, and Jochen Schiller.
\newblock A system for distributed event detection in wireless sensor networks.
\newblock In {\em Proceedings of the 9th ACM/IEEE International Conference on
  Information Processing in Sensor Networks - IPSN}, 2010.

\bibitem[ZHM{\etalchar{+}}12]{zhang12_statis_based_outlier_detec_wirel_sensor_networ}
Y.~Zhang, N.A.S. Hamm, N.~Meratnia, A.~Stein, M.~van~de Voort, and P.J.M.
  Havinga.
\newblock Statistics-based outlier detection for wireless sensor networks.
\newblock {\em International Journal of Geographical Information Science},
  26(8):1373--1392, 2012.

\bibitem[ZMH09]{zhang09_adapt_onlin_one_class_suppor}
Yang Zhang, Nirvana Meratnia, and Paul Havinga.
\newblock Adaptive and online one-class support vector machine-based outlier
  detection techniques for wireless sensor networks.
\newblock In {\em International Conference on Advanced Information Networking
  and Applications Workshops}, 2009.

\bibitem[ZR07]{zoumboulakis07_escal}
Michael Zoumboulakis and George Roussos.
\newblock {\em Escalation: Complex Event Detection in Wireless Sensor
  Networks}, pages 270--285.
\newblock Lecture Notes in Computer Science. Springer Science + Business Media,
  2007.

\end{thebibliography}
\end{document}